\documentclass[%
 amsmath,amssymb,
 aps,
]{revtex4-2}

\usepackage{graphicx}
\usepackage{epstopdf, epsfig}
\usepackage{amsfonts}
\usepackage{float}
\usepackage{subfig}
\usepackage{color}

\newcommand{\ks}{\textcolor{black}} 
\def\We{{\it We}}
\def\eff{{\it eff}}

\begin{document}


\title{Droplet breakup and size distribution in an airstream - effect of inertia}

\author{Someshwar Sanjay Ade$^1$, Pavan Kumar Kirar$^2$, Lakshmana Dora Chandrala$^3$\footnote{lchandrala@mae.iith.ac.in } and Kirti Chandra Sahu$^2$\footnote{ksahu@che.iith.ac.in}}
\affiliation{ 
$^1$Center for Interdisciplinary Program, Indian Institute of Technology Hyderabad, Kandi - 502 284, Sangareddy, Telangana, India \\
$^2$Department of Chemical Engineering, Indian Institute of Technology Hyderabad, Kandi - 502 284, Sangareddy, Telangana, India\\
$^3$Department of Mechanical and Aerospace Engineering, Indian Institute of Technology Hyderabad, Kandi - 502 284, Sangareddy, Telangana, India}%

\date{\today}

\begin{abstract}
We experimentally investigate the morphology and breakup of a droplet as it descends freely from a height and encounters an airstream. The size distributions of the child droplets are analysed using high-speed shadowgraphy and in-line holography techniques. We found that a droplet falling from various heights exhibits shape oscillations due to the intricate interplay between inertia and surface tension forces, leading to significant variations in the radial deformation of the droplet, influencing the breakup dynamics under an identical airstream condition. Specifically, the droplet undergoes vibrational breakup when introduced at a location slightly above the air nozzle. In contrast, as the release height of the droplet increases, keeping the Weber number defined based on the velocity of the airstream fixed, a dynamic interplay between the inertia of the droplet and the aerodynamic flow field comes into play, resulting in a sequence of breakup modes transitioning from vibrational breakup to retracting bag breakup, bag breakup, bag-stamen, retracting bag-stamen breakup, and eventually returning to vibrational breakup. Our experiments also reveal that the size distribution resulting from retracting bag breakup primarily arises from rim and node fragmentation, leading to a bimodal distribution. In contrast, bag and bag-stamen breakups yield a tri-modal size distribution due to the combined contributions of bag, rim, and node breakup mechanisms. Furthermore, we utilize a theoretical model that incorporates the effective Weber number, considering different release heights. This model accurately predicts the size distribution of the child droplets resulting from the various breakup modes observed in our experiments.
\end{abstract}


\maketitle

\section{Introduction} \label{sec:intro}
Droplet fragmentation and the resulting size distribution of child droplets play a significant role in a wide range of natural phenomena and industrial applications, including rainfall estimation, combustion, surface coating, pharmaceutical production, disease transmission modelling, artificial rain technology, and many others \citep{villermaux2007fragmentation,jain2019secondary,villermaux2020fragmentation,hopfes2021secondary,raut2021microphysical,xu2022droplet,kant2022bags,traverso2023data,balla2020numerical}. For instance, meteorological radar systems rely on droplet size distribution (DSD) to precisely predict rainfall \citep{yau1996short,Villermaux2009single}. In agricultural applications, the distribution of droplet sizes and velocities plays a crucial role in ensuring precise delivery and effective retention of pesticides on the intended targets \citep{ellis1997effect}. 
In various combustion applications, such as gas turbines, liquid rocket engines, ramjets and scramjets, achieving efficient atomization and controlling the resulting droplet size distribution are critical and fundamental requirements \citep{lefebvre2017atomization}. Thus, several researchers have investigated the fragmentation process of a spherical droplet as it enters the potential core region within an airstream \citep{pilch1987use, guildenbecher2009secondary, suryaprakash2019secondary, soni2020deformation,kulkarni2023interdependence}. 

A droplet undergoes morphological changes due to the aerodynamic force during its early inflation stage when it interacts with an airstream. At the later stages, it experiences fragmentation caused by the Rayleigh-Taylor instability, Rayleigh-Plateau capillary instability \citep{taylor1963shape} and the nonlinear instability of liquid ligaments \citep{jackiw2021aerodynamic, jackiw2022prediction}. The dynamics of a droplet in an airstream are influenced by the interplay between aerodynamic and surface tension forces, characterized by the Weber number $(\We)$. This is defined as $\We \equiv \rho_a U^2 d_0/\sigma$, where $\rho_a$, $\sigma$, $U$ and $d_0$ denote the air density, interfacial tension, average velocity of the airstream and equivalent spherical diameter of the droplet, respectively. As the Weber number increases, the droplet exhibits various breakup modes, including vibrational, single-bag, bag-stamen, multi-bag, shear mode and catastrophic fragmentation.

In the case of vibrational breakup at low Weber numbers ($\We$), the droplet experiences shape oscillations at a specific frequency. Eventually, when the amplitude of these oscillations reaches a level comparable to the size of the droplet, fragmentation occurs. As the Weber number increases, the drop evolves into a single bag on the leeward side, surrounded by a thick liquid rim. The bag and rim breakups lead to the formation of tiny and slightly bigger droplets. This fragmentation process is known as the bag breakup phenomenon \citep{taylor1963shape,kulkarni2014bag,soni2020deformation}. The critical Weber number $(\We_{cr})$ for the transition from the vibrational to the bag breakup is approximately 12. The bag-stamen and multi-bag modes have similar morphologies to the bag breakup mode but with an additional stamen formation in the centre of the drop. This results in a large additional drop (bag-stamen) and several bags (multiple bags) during the breakup process. For intermediate Weber numbers ($28 \le \We \le 41$), \citet{cao2007new} observed the dual-bag breakup mode through shadowgraph and high-speed imaging. In the shear mode observed at higher Weber numbers, the edge of the droplet deflects downstream, causing the sheet to fracture into tiny droplets. For very high Weber numbers, droplet quickly explodes into a cluster of small fragments, resulting in catastrophic fragmentation. Recently, \citet{ade2023size, joshi2022droplet, boggavarapu2021secondary} conducted experimental investigations and observed bag, bag-stamen, dual-bag, and multi-bag breakup modes for different Weber numbers. \citet{kirar2022experimental, ade2022droplet} explored the fragmentation of a droplet exposed to a swirl airstream with a fixed Weber number using a shadowgraphy technique. Their investigation led to the finding of a new breakup mechanism known as the retracting bag breakup mode for intermediate swirl strengths. In this mode, the ligaments experience stretching in opposite directions due to the swirling airstream, resulting in the droplet undergoing capillary instability and ultimately breaking apart.  

Several experimental techniques, such as planar methods, imaging techniques, laser-induced fluorescence, and femtosecond pulsed light sources, have been used to investigate the droplet breakup phenomenon. In a comprehensive review, \citet{tropea2011optical} highlighted the advancements made in measuring particle size, temperature, and composition with these techniques. \citet{boggavarapu2021secondary} explored the droplet size distribution associated with different breakup modes using the particle/droplet image analysis (PDIA) method. They found that the bag and bag-stamen breakups exhibited a tri-modal size distribution, while the dual-bag and multi-bag breakups produced a bi-modal droplet size distribution. Digital in-line holography has recently emerged as a powerful tool for estimating droplet size distribution \citep{guildenbecher2017characterization,shao2020machine,radhakrishna2021experimental,essaidi2021aerodynamic,li2022secondary,ade2022droplet,ade2023size}. \citet{radhakrishna2021experimental} investigated the effect of the Weber number on droplet fragmentation at high Ohnesorge numbers and examined various breakup modes using the digital in-line holography technique. Recently, \citet{ade2023size,ade2022droplet} employed the digital in-line holography technique to investigate droplet fragmentation in straight and swirl airstreams. \citet{ade2023size} explored the droplet size distribution for different Weber numbers undergoing single-bag and multi-bag fragmentations. They demonstrated that despite having six distinct breakup mechanisms, the dual-bag breakup exhibits a bi-modal distribution, in contrast to the single-bag breakup, which undergoes a tri-modal distribution. Moreover, they found that the analytical model proposed by \citet{jackiw2022prediction} accurately predicts the DSD for a wide range of Weber numbers.

In all the aforementioned studies, a droplet with a zero initial velocity was introduced into the potential core region of the airstream generated by a nozzle. This approach neglects the inertia of the droplet as it falls freely and interacts with the surrounding airstream. It is well known that raindrops falling from the cloud do not even attend terminal velocity and undergo topological changes due to the associated microphysical processes, such as phase change, collision-coalescence and breakup \citep{montero2009all}. Also, in various other applications, the droplet exhibits inertia and retains a non-zero downward velocity when encountering an airstream. Thus, the primary objective of the present study is to investigate how droplet inertia impacts its morphology and breakup behavior when subjected to an airstream using the high-speed shadowgraphy technique. When a droplet is dispensed from a height, it experiences an accelerating phase, and its subsequent dynamics result from a complex interplay between its weight, the aerodynamic force of the airstream, and the inertia of the droplet. Interestingly, we observe that droplets falling from different heights exhibit distinct breakup modes even when encountering the same fixed airstream (constant Weber number). Using the in-line holography technique, we also investigate the size distribution of the child droplets associated with these breakup modes. Furthermore, we incorporate an effective Weber number of the droplet on the analytical model developed by \citet{jackiw2022prediction}, which satisfactorily predicts the size distribution observed in our experiments. These aspects remain largely unexplored, making our study unique and significant in advancing knowledge in the field.

The rest of the paper is organised as follows. The experimental setup and procedure are discussed in \S\ref{sec:expt}. The impact of dispensing height on droplet fragmentation and the resulting size distribution is discussed in \S\ref{sec:dis}. Additionally, in this section, we present a theoretical model in order to predict the size distributions of satellite droplets resulting from various breakup modes at different heights. Finally, the concluding remarks can be found in \S\ref{sec:conc}.

\section{Experimental procedure}\label{sec:expt}

We employ shadowgraphy and digital in-line holography techniques to record the morphology of a droplet as it is released from varying heights and to estimate the size distribution of the child droplets resulting from the different breakup modes influenced by an airstream. An air nozzle with an inner diameter of $D=18$ mm is connected to a digital mass flow controller (model: MCR-500SLPM-D/CM, Make: Alicat Scientific, Inc., USA) and an air compressor. The airflow rate through the nozzle is maintained constant at 228 litres per minute. To investigate the fragmentation process, we generate water droplets with an initial diameter of $d_0=3.2 \pm 0.07$ mm using a 20-gauge dispensing needle with a diameter of $0.908$ mm. A Cartesian coordinate system $(x, y, z)$ is employed, with the nozzle center as the origin to analyze the droplet dynamics and breakup phenomenon. The normalised height of the dispensing needle ($h/D$) is varied from 0.6 to 11.2 to investigate the effect of the inertia of the droplet on the fragmentation process, while the other coordinates of the dispensing needle, $x_{d}/D$ and $z_{d}/D$ are kept fixed at 0.36 and 0, respectively. A Weber number of $\We=12.1$ is maintained in all the experiments, which corresponds to a scenario when a droplet undergoes single-bag fragmentation when it is introduced into the potential core region of the airstream \citep{taylor1963shape,kulkarni2014bag,soni2020deformation}.

\begin{figure}
\centering
\includegraphics[width=0.6\textwidth]{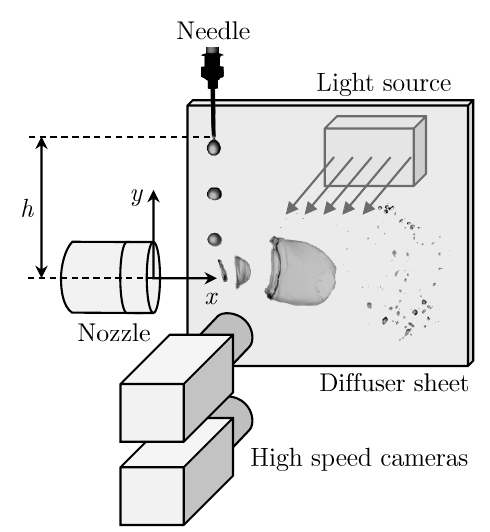}
\caption{Schematic diagram of the experimental set-up for shadowgraphy. It consists of two high-speed cameras with a diffuser sheet and two light sources, an air nozzle, and a droplet dispensing needle to generate the droplet of constant size. The top camera captures the motion of the droplet from the dispensing needle to the air nozzle. The bottom camera captures the droplet deformation and fragmentation in the airstream.}
\label{fig1a}
\end{figure}

\begin{figure}
\centering
\includegraphics[width=0.9\textwidth]{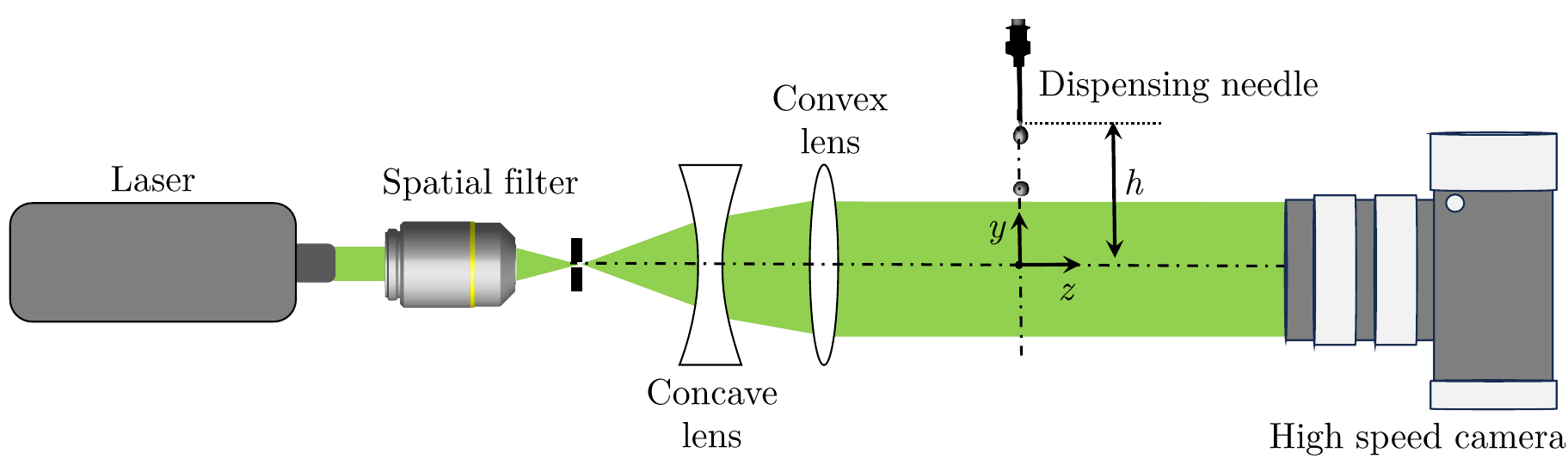}
\caption{Schematic diagram of the digital in-line holography experimental set-up (side view).}
\label{fig1b}
\end{figure}

Figure \ref{fig1a} illustrates the schematic diagram for shadowgraphy. This setup comprises two high-speed imaging systems equipped with high-speed cameras synchronized using a digital delay generator (model: 610036, Make: TSI, USA), a diffuser sheet, and light sources. The top camera records the droplet dynamics from the dispensing needle to the air nozzle, while the bottom camera records the droplet dynamics in the airstream. The high-power light source (model: MultiLED QT, Make: GSVITEC, Germany) and a uniform diffuser sheet are used to illuminate the background. The high-speed cameras are configured with the following settings: a frame rate of 1400 frames per second, exposure duration of 1 $\mu$s, and a resolution of $1596 \times 3361$ pixels. These cameras have fixed focal length lenses of 135 mm, resulting in a spatial resolution of 68.12 $\mu$m/pixel.

Figure \ref{fig1b} shows the schematic representation of the in-line holography setup (side view). This setup comprises several key components, including a continuous wave laser (model: SDL-532-100T, Shanghai Dream Lasers Technology Co. Ltd.), a spatial filter system, and collimating optics (Holmarc Opto-Mechatronics Ltd.), and a high-speed camera (model: Phantom VEO 640L, Vision Research, USA). The high-speed camera is positioned at a distance of $z=500$ mm, as depicted in figure \ref{fig1b}. The continuous wave laser generates a coherent beam with a wavelength of 532 nm. The initial step involves passing this laser beam through a spatial filter consisting of an infinity-corrected plan achromatic objective boasting 20X magnification ( Holmarc Opto-Mechatronics Ltd.) and a 15 $\mu$m pin-hole. The purpose of the spatial filter is to efficiently remove unwanted noise, allowing only the clean portion of the beam to pass through the pin-hole. After the spatial filtering process, the beam undergoes expansion through a plano-concave lens, followed by collimation achieved with a plano-convex lens. The resultant collimated beam is then directed toward the droplet field.  The interference of scattered light emanates from the droplets, and the unscattered light generates a hologram on the sensor. The holograms are captured using the high-speed camera at a frame rate of 1400 frames per second (fps) with 1 $\mu$s exposure time, resulting in images of dimensions $1664 \times 1600$ pixels. The recorded holograms have a spatial resolution of 40.70 $\mu$m/pixel.

A detailed illustration of the digital in-line holography and the associated post-processing method can be found in Refs. \cite{ade2023size,ade2022droplet,ade2024application}. In the following section, the dimensionless time, $\tau=Ut\sqrt{\rho_a/\rho_w}/d_0$ is used to describe the results, such that $\tau = 0$ indicates the instant the droplet enters the aerodynamic field. Here, $t$ denotes time in second and $\rho_w$ represents the density of water (998 kg m$^{-3}$).

\section{Results and discussion}\label{sec:dis}

We investigate the effect of the inertia of a droplet, released from different heights and exposed to a horizontal airstream, on its morphology, breakup behaviour and the resultant size distribution of the child droplets. To focus exclusively on the impact of dispensed height, we maintain a constant Weber number ($\We =12.1$) in this study by keeping the diameter of the parent drop ($d_0 =3.2 \pm 0.07$ mm) and the average velocity of the airstream $(U)$ fixed. 

\subsection{Drop morphology} \label{sec:morp}

\begin{figure}
\centering
\includegraphics[width=0.9\textwidth]{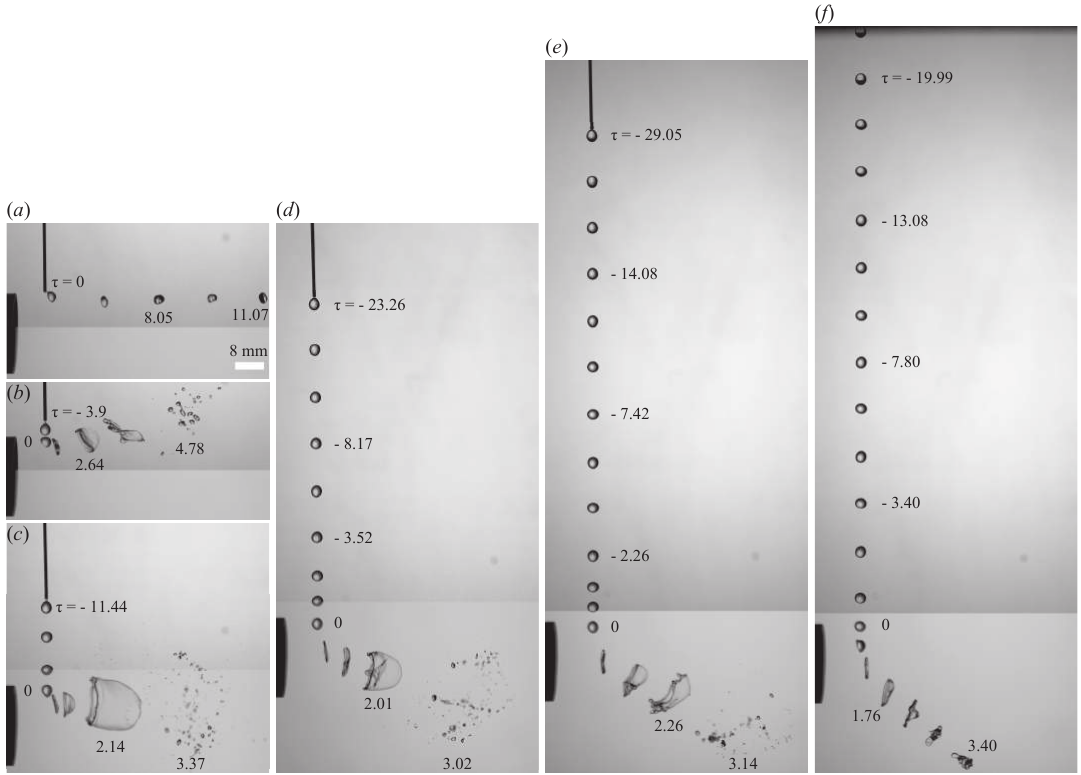}
\caption{Temporal evolution of the droplet breakup dynamics for (a) $h/D = 0.6$ (vibrational mode), (b) $h/D = 0.8$ (retracting bag breakup), (c) $h/D = 1.9$ (bag breakup), (d) $h/D = 5.6$ (bag-stamen breakup), (e) $h/D = 8.4$ (retracting bag-stamen breakup) and (f) $h/D = 11.2$ (vibrational breakup) at $\We = 12.1$. The dimensionless time ($\tau=Ut\sqrt{\rho_a/\rho_w}/d_0$) is mentioned in each panel, wherein $U$ is the free stream velocity, $t$ is the physical time, $\rho_a$ is the density of air, $\rho_w$ is the density of the water and $d_0$ is the droplet diameter. The instant, $\tau=0$ represents the time when the droplet starts entering into the aerodynamic field. The droplet breakup phenomena associated with panels (a-f) are provided as supplementary movies (1–6), respectively.}
\label{fig2}
\end{figure}

\begin{figure}
\centering
\includegraphics[width=0.7\textwidth]{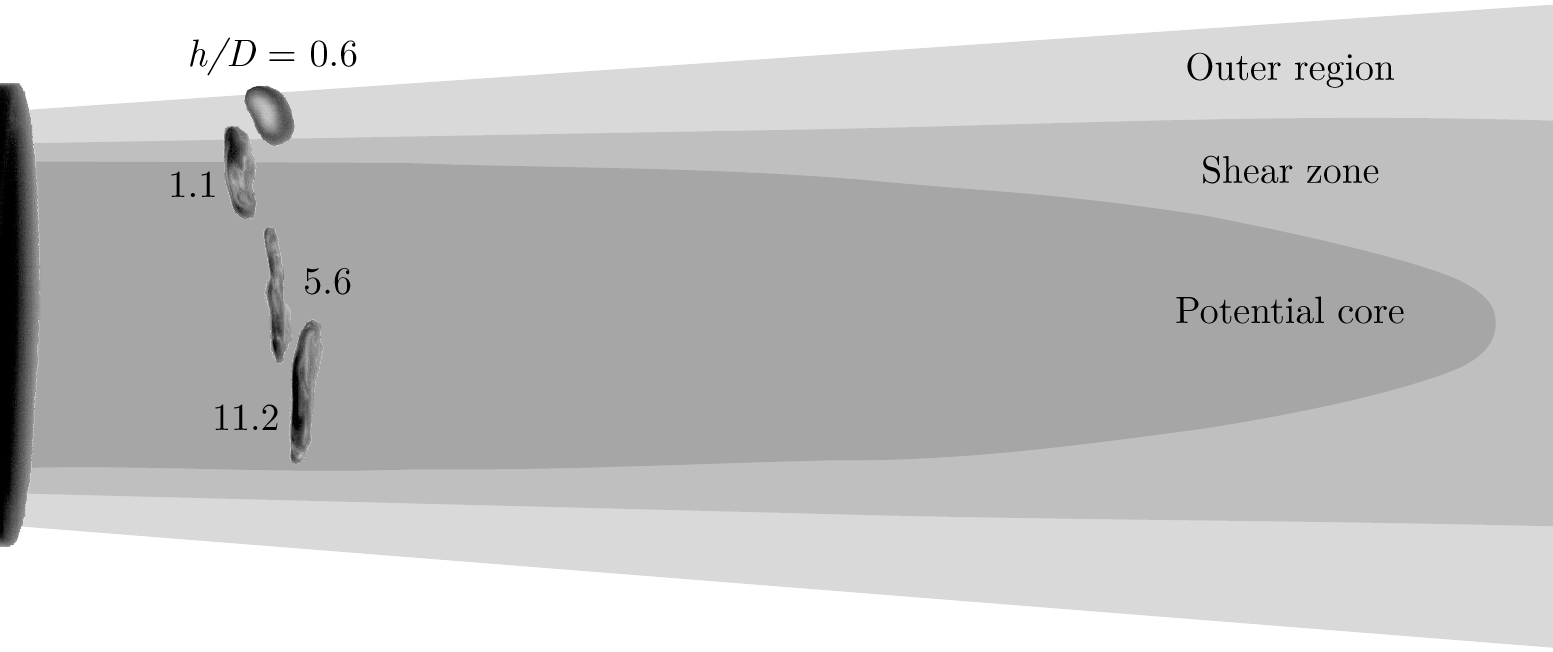}
\caption{Schematic representation of the relative positions of the droplet in various zones of the airstream during the fragmentation process for different dispensing heights, $h/D$.}
\label{fig2b}
\end{figure}

Figure \ref{fig2} depicts the droplet dynamics for different values of the normalised released height $(h/D)$ and subjected to an airstream in a cross-flow configuration. Based on the stereo-particle image velocimetry (PIV) measurements of the airstream, as detailed \citet{kirar2022experimental}, in figure \ref{fig2b}, we illustrate a schematic representation of the relative positions of droplets in various flow zones during the fragmentation process for different dispensing heights, $h/D$. The airstream can be characterised by the potential core, shear and outer regions as depicted in figure \ref{fig2b}. Figure \ref{fig2}(a) shows the temporal evolution of drop when it is released into the aerodynamic field from a height of $h/D = 0.6$ (slightly above the air nozzle). Due to its minimal potential energy and reduced inertia, the droplet cannot penetrate the airstream and remains positioned at the outer periphery of the airflow. In this outer region of the airstream, the aerodynamic force lacks the necessary intensity to trigger fragmentation. Consequently, the droplet exhibits a vibrational mode. Figure \ref{fig2}(b) depicts the droplet dynamics for $h/D = 0.8$. In this scenario, the droplet possesses sufficient inertia to traverse the outer region of the aerodynamic flow field, and it enters the shear zone. In the shear zone, the strong aerodynamic forces lead to the deformation of the droplet from a sphere to a tilted disk, resulting from the pronounced velocity gradients present within the shear zone. Subsequently, the aerodynamic influence induces the formation of a liquid sheet from the tilted disk. This sheet retracts in a direction opposing the prevailing airflow, driven by a negative pressure gradient, and the droplet undergoes fragmentation due to capillary instability \citep{taylor1963shape}. A similar breakup pattern was observed by \citet{kirar2022experimental} in a swirling airflow, which they referred to as ``retracting bag breakup''. Increasing the droplet released height further ($h/D=1.9$), it can be seen in figure \ref{fig2}(c) that the droplet's inertia becomes sufficient to overcome the resistance posed by both the outer region and the shear zone of the aerodynamic field. As a result, approximately half of the droplet takes position within the shear zone, while the remaining half occupies the potential core region of the aerodynamic field. In both these regions, the aerodynamic force is strong enough to counterbalance the surface tension force of the droplet. This interaction causes the droplet to change its shape from a sphere to a disk. Due to the Rayleigh-Taylor instability, air infiltrates the centre part of the formed disk, creating a central bag-like structure and a toroidal rim encircling its periphery. Subsequently, the tip of the bag ruptures due to the exerted aerodynamic forces, while the rim disintegrates due to capillary instability.

It can be seen in figure \ref{fig2}(d) and (e) that the droplet dispensed from $h/D=5.6$ and 8.4 gets sufficient inertia and downward velocity to traverse through the potential core region of the airstream while it undergoes a rapid change in topology from a spherical shape to a disk shape. For  $h/D=5.6$ (figure \ref{fig2}d), the outer edge of the disk-shaped drop is elongated into a thin sheet due to drag forces. The central region of the disk is subject to Rayleigh-Taylor instability, which occurs because the lighter air is accelerated against the heavier liquid, leading to the formation of a bag-like structure. The Rayleigh-Taylor instability is predominant at the interface where the liquid disk meets the air, causing the central region to expand outward and form a bag. The stamen that appears at the center of the bag is due to the localized thinning and elongation of the liquid as it is pulled by both aerodynamic forces and surface tension effects, accompanied by a toroidal rim surrounding its periphery $( \tau=2.01 )$. The bag eventually ruptures as the dominant aerodynamic forces exceed the surface tension. In contrast, the toroidal rim and stamen fragment due to capillary instability $( \tau=3.02 )$ at the lower part of the air nozzle. This entire process is termed the bag-stamen breakup phenomenon. For $h/D=8.4$ (figure \ref{fig2}e), the higher downward velocity enables the deformed disk to cross the potential core region, with part of the disk entering the lower shear zone of the airstream. Here, the outer edge of the disk experiences elongation into a thin sheet due to drag forces, while the central portion undergoes Rayleigh-Taylor instability, extending outward to form a bag. This results in a distinct bag-stamen pattern within the droplet. The bag-stamen structure inclines due to the large velocity gradient within the lower shear layer. During the subsequent period, the bag retracts, due to a negative pressure gradient, and the droplet fragments due to capillary instability. This sequence of events is termed the retracting bag-stamen breakup phenomenon. For the highest value of $h/D$ value considered in this study ($h/D=11.2$), the droplet possesses significant potential energy and greater inertia to migrate into the potential core with a higher downward velocity. The droplet undergoes morphological change from sphere to disk quickly while the inertia effect is sufficiently strong to cross the potential core and lower the shear layer. As a result, the droplet enters the outer region of the lower portion of the flow field and undergoes vibrational breakup. The characteristic time scales for the Rayleigh-Taylor and capillary instabilities align with the observed time scales of the breakup process. The Rayleigh-Taylor instability develops rapidly as the droplet enters the high-speed airstream, forming the bag within milliseconds. The subsequent breakup of the bag, toroidal rim, and stamen due to capillary instability also occurs on a similarly short-time scale, driven by the competing forces of surface tension and aerodynamic drag.

\begin{figure}
\centering
{\large (a)}  \\
\includegraphics[width=0.9\textwidth]{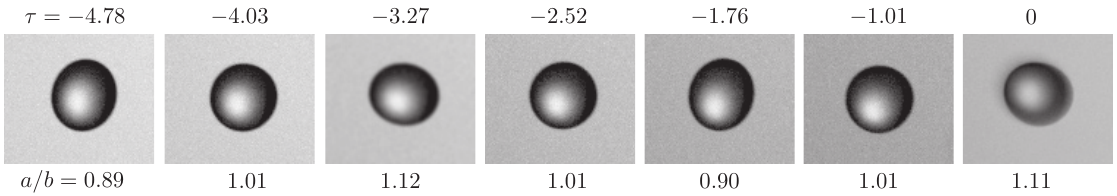}\\
{\large (b)}\\
\includegraphics[height=0.31\textwidth]{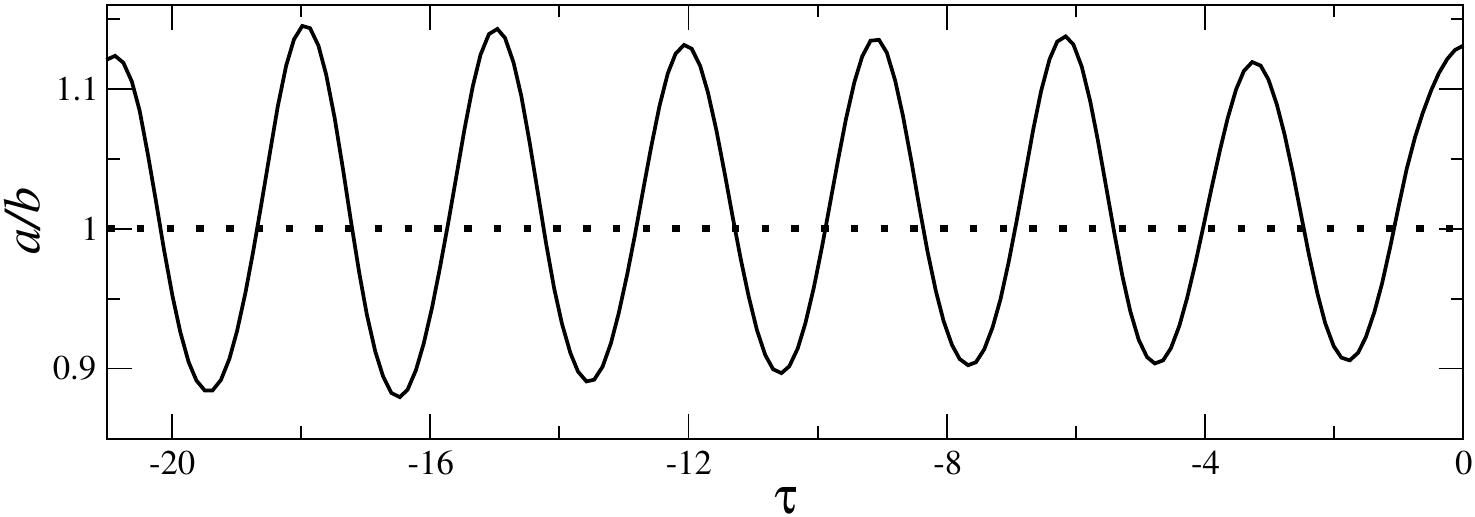}
\caption{Temporal variation of (a) the shape of the drop and (b) the aspect ratio $(a/b)$ during its free fall for $h/D = 11.2$. The dimensionless time, $\tau$ is mentioned at the top of each panel. The value of the aspect ratio $(a/b)$ is mentioned at the bottom of each drop shape in panel (a). Here, $a$ and $b$ are the diameters of the droplet in $x$ and $y$ directions, respectively.}
\label{fig7}
\end{figure}

\begin{figure}
\centering
\includegraphics[width=0.95\textwidth]{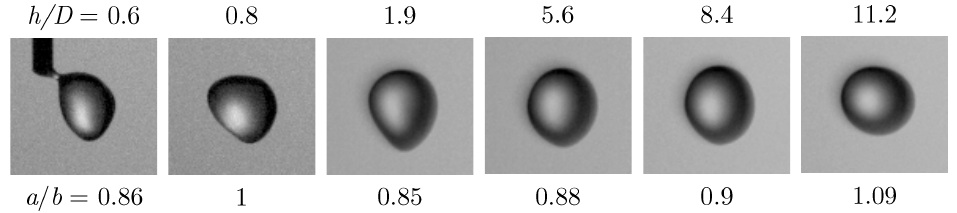}
\caption{Shape of the drop at the onset of its entrance into the airstream (at $\tau = 0$) for different values of $h/D$. The corresponding aspect ratio $(a/b)$ is indicated at the bottom of each panel.}
\label{fig3b}
\end{figure}

A falling drop exhibits natural oscillations in its shape, transitioning between oblate, spherical and prolate shapes as it descends from a height due to the competition between inertia and surface tension forces. It detects the shape of the droplet as it enters the potential core region, which also influences the drop breakup phenomenon. Thus, in order to gain better insights, we investigate the shape oscillation of the drop during its free fall after release from the needle. The frequency $(f)$ of the natural oscillations of a freely suspended water drop derived using a linear theory in the inviscid limit is given by \citep{rayleigh1879capillary,nelson1972oscillation}:
\begin{equation}
f = \left[ n (n-1) (n+2)  \left ( {2 \sigma \over \pi^2 \rho {d_0}^3} \right) \right]^{1/2},   \label{Tp_Theory}
\end{equation}
where $n$ denotes the oscillation mode, with the fundamental mode corresponding to $n=2$. The time period $(T_p)$ of the natural oscillations of a drop is given by $1/f$. In figure \ref{fig7}, we investigate the temporal variation of the drop shape and the aspect ratio $(a/b)$ during its free fall for $h/D = 11.2$. Here, $a$ and $b$ are the droplet diameters in $x$ and $y$ directions, respectively, such that $a/b > 1$, $a/b = 1$ and $a/b < 1$ represent oblate, spherical and prolate shapes, respectively. It can be seen that the falling drop exhibits periodic shape oscillations about the spherical shape. These symmetric and periodic oscillations have also been reported in earlier investigations in the context of a falling nonspherical droplet in the air \citep{agrawal2017nonspherical,balla2019shape,agrawal2020experimental}. Inspection of figure \ref{fig7} reveals that the dimensionless time period of oscillations is $\approx 2.97$, which closely matches with the theoretical prediction of the dimensionless time period obtained using Eq. (\ref{Tp_Theory}) for the fundamental mode ($(1/f) U \sqrt{\rho_a/\rho_w}/d_0 = 2.73$). Note that the time period of the shape oscillations of the primary droplet remains unchanged for different release heights, as it solely depends on the liquid properties and the initial size of the droplet (Eq. \ref{Tp_Theory}). Our experiments also observe the same time period regardless of the height from which the droplet is released (not shown). However, as expected, the number of shape oscillation cycles decreases with the decrease in the release height. Figure \ref{fig3b} shows the shape of the droplet at the instant when it enters the aerodynamic field (represented by $\tau=0$) for different values of $h/D$. As expected, it can be observed that there is a substantial influence of the descent height, and consequently its inertia of the drop, on both its shape and aspect ratio before encountering the airflow. This, in turn, influences the ensuing droplet breakup phenomenon.

\begin{figure}
\centering
\includegraphics[width=0.9\textwidth]{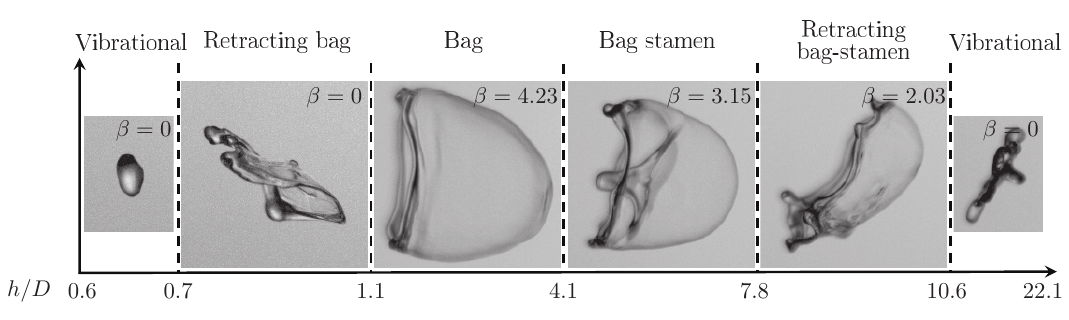}
\caption{Shape of the drop at the onset of various breakup modes for different values of $h/D$. The value of the normalised bag inflation, $\beta = l/d_{0}$, is mentioned at the top right corner of each panel. Here, $l$ denotes the length of the bag in streamwise direction.}
\label{fig4}
\end{figure}

Figure \ref{fig4} illustrates the morphology of the droplet for various breakup modes corresponding to different values of $h/D$. The droplet exhibits a vibrational mode for $h/D = 0.6$ to $0.7$. As discussed above, this phenomenon can be attributed to the position of the drop within the outer region of the aerodynamic field. In the range of $h/D = 0.7$ to $1.1$, the drop experiences a retracting bag breakup due to its presence within the shear region of the aerodynamic field. For $h/D = 1.1$ to $4.1$, the drop
undergoes a bag breakup mode as a result of the combined influences of the shear region and the potential core within the aerodynamic field. As we increase the value of $h/D$ further, between $h/D=4.1$ to $7.8$, the droplet enters a distinct bag-stamen breakup phenomenon driven by its position within the potential core of the aerodynamic field. Subsequently, for $h/D = 7.8$ to $10.6$, the droplet transitions to a retracting bag-stamen breakup due to its presence within the potential core and the lower shear layer of the aerodynamic field. For $h/D = 10.6$ to $22.1$, the droplet again displays vibrational breakup mode attributed to its exposure to the lower outer region of the aerodynamic field. A close inspection of figure \ref{fig4} also reveals that the normalised bag inflation in the streamwise direction, defined as $\beta = l/d_{0}$, increases and then decreases as we increase the value of $h/D$.

\subsection{Trajectory}

\begin{figure}
\centering
\includegraphics[width=0.55\textwidth]{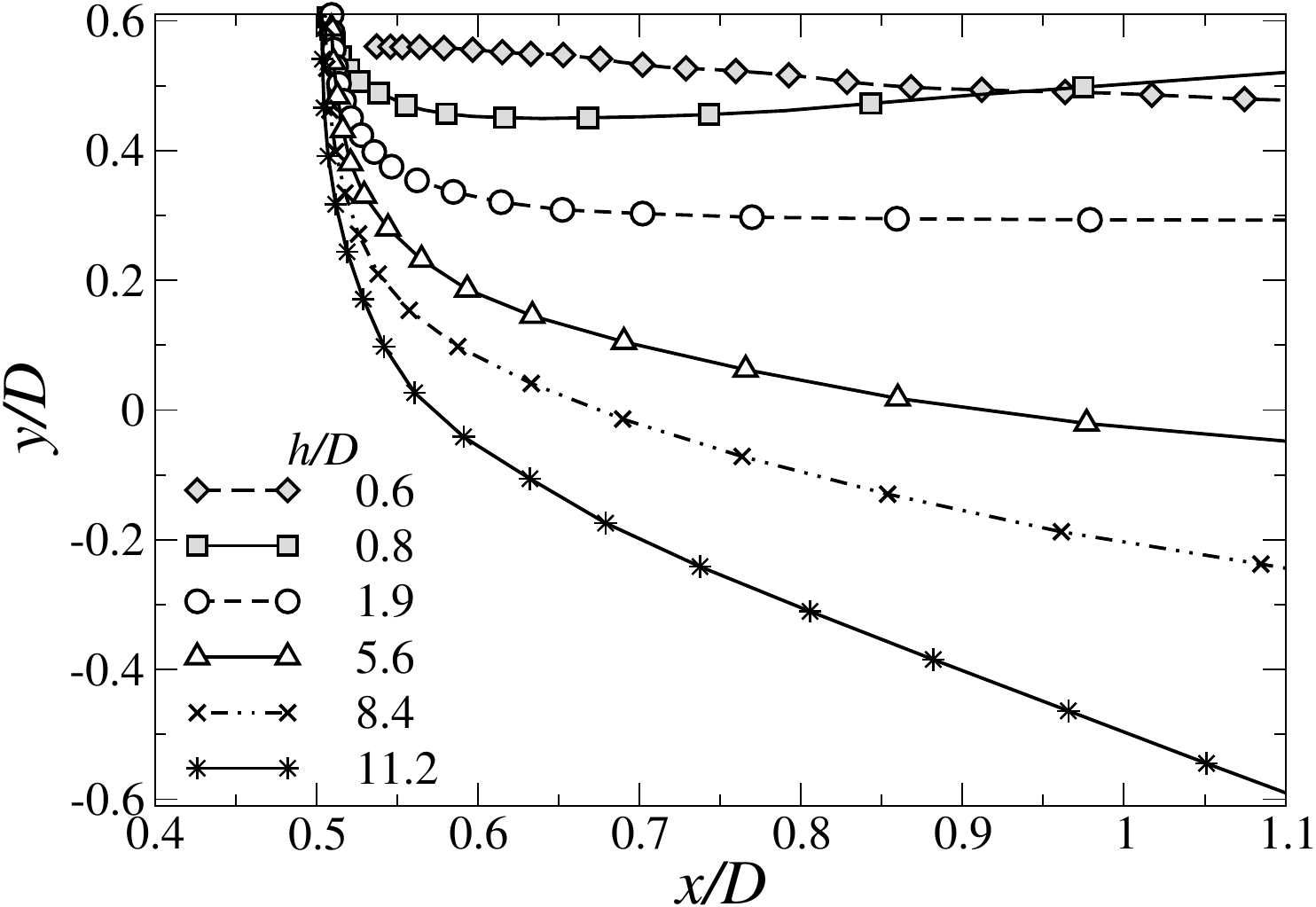}
\caption{Trajectories of the droplet for $h/D = 0.6$ (vibrational mode), $h/D = 0.8$ (retracting bag breakup), $h/D = 1.9$ (bag breakup), $h/D = 5.6$ (bag-stamen breakup), $h/D = 8.4$ (retracting bag-stamen breakup) and $h/D = 11.2$ (vibrational breakup) at $\We = 12.1$. Note that the position of the droplet is determined based on its centroid. The symbols (squares, circle, triangles, etc.) represent measured data points.}
\label{trajectory_figure}
\end{figure}

Figure \ref{trajectory_figure} depicts the droplet trajectories in the aerodynamic field before the onset of fragmentation for different values of $h/D$. For $h/D = 0.6$, the droplet possesses minimal inertia, restricting its vertical descent to the outermost region of the aerodynamic field. In this outer region, the droplet travels horizontally along the predominant airstream direction and undergoes a vibrational mode. For $h/D = 0.8$, the droplet has sufficient inertia to cross the outer region and migrates vertically downward up to the shear zone of the aerodynamic field. In this case, the droplet travels in both horizontal and vertical upward directions in the shear region due to large velocity gradients. As a result, the droplet exhibits the retracting bag breakup mode. For $h/D = 1.9$, the droplet has high inertia to migrate vertically downward and get exposed to both the shear and potential core zone of the aerodynamic field. In this case, the droplet travels horizontally along the airstream and exhibits bag breakup mode. For $h/D = 5.6$, the droplet has higher inertia to migrate vertically downward till the potential core region of the aerodynamic field. In this region, the droplet travels horizontally due to the drag of the airstream and vertically downward due to the inertia effect while undergoing a bag-stamen fragmentation. For $h/D = 8.4$, the droplet possesses more substantial inertia to travel vertically downward and get exposed to the potential core and lower shear region of the aerodynamic field. In this case, the droplet exhibits retracting bag-stamen breakup and generates a trajectory similar to the bag-stamen mode. For $h/D = 11.2$, the droplet travels vertically downward with a high velocity and first enters the potential core. Later, due to the high inertia effect, the droplet further crosses the potential and lower shear zone and enters the lower outer region of the flow field. In this case, the droplet trajectory follows less migration in the predominant airstream direction and more travel vertically downward due to a more substantial inertia effect. It is important to note that for the vibrational, retracting bag, and bag breakup modes, the trajectory of the droplet ends above the central axis of the nozzle within the aerodynamic field. In contrast, for the bag stamen and retracting bag-stamen modes, the trajectory of the droplet concludes below the central axis of the nozzle. 

\subsection{Effective Weber number}

\begin{figure}
\centering
\includegraphics[width=0.55\textwidth]{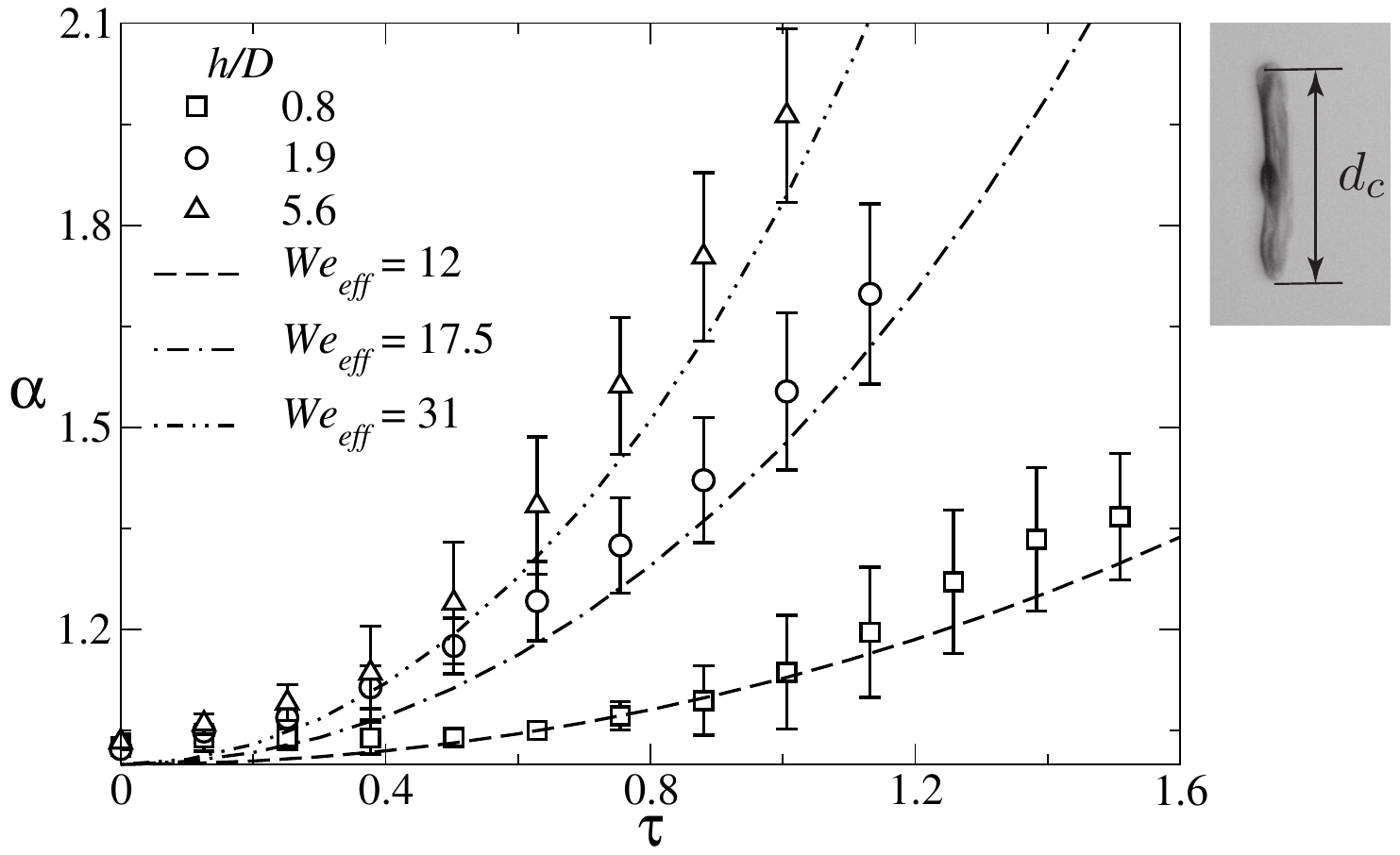} 
\caption{Temporal evolution of radial deformation of the drop, $\alpha=d_c/d_0$ for different values of $h/D$. Here, $d_c$ denotes the diameter of the disk. The symbols represent the experimental results with an error-bar associated with the standard deviation from three repetitions. The dashed lines represent the theoretical results.}
\label{fig6}
\end{figure}

As discussed in \S\ref{sec:morp}, the deformation of the droplet undergoes significant variations during its interaction with the airstream when released from different heights, primarily due to changes in inertia, even though the Weber number for the airstream is maintained fixed ($\We=12.1$). Due to the intricate interplay between the inertia of the droplet and the aerodynamic flow field, the Weber number $(\We)$ calculated based on the air velocity at the nozzle exit fails to adequately describe the various droplet breakup modes associated with different release heights. Our experimental results demonstrate that the release height $(h/D)$ has a profound impact on the radial deformation of the droplet, which, in turn, influences both the breakup morphology and the resulting sizes of the child droplets. Consequently, it becomes imperative to employ an effective Weber number $(\We_{\eff})$ that considers the effect of the radial deformation of the droplet in the analytical model (discussed in the following section) to estimate the size distribution of the resulting child droplets \citep{kulkarni2014bag,kirar2022experimental}. Figure \ref{fig6} depicts the temporal evolution of the radial deformation of the droplet, denoted as $\alpha=d_{c}/d_0$, for different values of $h/D$. Here, $d_{c}$ represents the diameter of the disk-shaped droplet, which is illustrated in the inset of figure \ref{fig6}. For $h/D = 0.8$, the droplet experiences minimal radial deformation, primarily due to its residence in the shear zone of the airstream, where it encounters weak aerodynamic forces. For $h/D=1.9$, the droplet displays more substantial radial deformation as it interacts with both the shear and potential core regions of the airstream, subjecting it to stronger aerodynamic forces. For $h/D=5.6$, the droplet undergoes the most extensive radial deformation as it migrates into the potential core of the airflow, exposing it to the strongest aerodynamic forces. 

Considering the radial deformation of the drop for various $h/D$ values, we derive an expression for the effective Weber number  $(\We_{\eff})$ based on the analyses presented in  Refs. \cite{Villermaux2009single,kulkarni2014bag,kirar2022experimental}. We start our analysis by examining the aerodynamic field around the drops exposed to the airstream  \cite{Villermaux2009single}. Let $U = (u_r, u_y)$ represent the axisymmetric velocity field near the drop. The continuity equation for the airflow around the drop, assuming steady, inviscid, and incompressible flow, is given by:
\begin{equation}
r\frac{\partial u_{y}}{\partial y}+\frac{\partial (ru_{r})}{\partial r}=0.
\label{continuity_equation}
\end{equation}
Due to the aerodynamic pressure imbalance between the aft and rear sides of the drop, the drop deforms into a flat disk aligned with the airstream direction. This flow can be modeled as a stagnation point flow, where $u_y = -\gamma y$, with $ \gamma$ being the stretching rate, $y$ the distance from the drop surface on the front side, and $r$ a direction pointing radially outward. Thus, $u_{r}=r\gamma /2$. The pressure field $P_{a}(r,y)$ around the droplet can be obtained from the steady, inviscid, and incompressible momentum equations in the $r$ and $y$ directions, which are given by 
\begin{equation}
\rho _{a}u_{r}\frac{\partial u_{r}}{\partial r}=-\frac{\partial P_{a}}{\partial r} ~~ \textrm{and}~~
\rho _{a}u_{y}\frac{\partial u_{y}}{\partial y}=-\frac{\partial P_{a}}{\partial y}.
\label{r_and_y_momentum_equation}
\end{equation}
By substituting the expressions of $u_{y}$ and $u_{r}$ into the above equations and integrating the resultant equations, we obtain 
\begin{equation}
P_{a}(r,y)=\frac{\rho _{a}U^{2}}{2}-\frac{\rho _{a}r^{2}\gamma ^{2}}{8}+\frac{\rho _{a}y^{2}\gamma ^{2}}{2}.
\label{air_pressure_equation}
\end{equation}
Therefore, the pressure of air at the drop surface is
\begin{equation}
P_{a}(r,y=0)=\frac{\rho _{a}U^{2}}{2}-\frac{\rho _{a}r^{2}\gamma ^{2}}{8}.
\label{air_pressure_equation_at_y=0}
\end{equation}
Now, the pressure inside the disk-shaped droplet ($P_{l}$) can be evaluated as:
\begin{equation}
P_{l}=\frac{\rho _{a}U^{2}}{2}-\frac{\rho _{a}R(t)^{2}\gamma ^{2}}{8}+\frac{2\sigma }{h(t)}.
\label{drop_pressure_equation}
\end{equation}
Here, $R$ represents the radius of the disk-shaped droplet, $\sigma$ denotes for the surface tension at the liquid-gas interface, and $h$ denotes the thickness of the disk-shaped droplet. We employ the continuity equation for inviscid and incompressible fluid in order to evaluate the velocity field $(u_{l})$ inside the droplet, which is given by 
\begin{equation}
r\frac{\partial h}{\partial t}+\frac{\partial (ru_{l}h)}{\partial r}=0.
\label{drop_velocity_continuity}
\end{equation}
Now, substituting $h(t) = {d_{o}^3/6R^2}$ (obtained using the global mass conservation) into the above equation, and integrating the resultant equation yields
\begin{equation}
u_{l}=\frac{r}{R}\frac{\mathrm{d} R}{\mathrm{d} t}.
\label{drop_velocity_field}
\end{equation}
The momentum equation for the droplet is given by
\begin{equation}
\rho _{l}\left ( \frac{\partial u_{l}}{\partial t}+u_{l}\frac{\partial u_{l}}{\partial r} \right )=-\frac{\partial P_{l}}{\partial r}
\label{drop_momentum_equation}
\end{equation}
Non-dimensionalizing and integrating Eq. (\ref{drop_momentum_equation}) from $r=0$ to $r=R$ and using Eq. (\ref{drop_pressure_equation}), we obtain
\begin{equation}
\frac{\mathrm{d}^{2}\alpha  }{\mathrm{d} \tau ^{2}}-\left ( \frac{f^{2}}{4}-\frac{24}{We} \right )=0,
\label{diff_equation}
\end{equation}
where $\alpha (t)=2R/d_{0}=d_{c}/d_{0}$, and $f=\gamma d_{0}/U$ denotes the stretching factor. Now, solving Eq. (\ref{diff_equation}) using the initial conditions, given by $\alpha (0)=1$ and $\mathrm{d\alpha }/\mathrm{d \tau }=0$, we get \cite{kirar2022experimental}\begin{equation}
\alpha =\textrm{cosh}\left ( \tau \sqrt{\frac{f^{2}}{4}-\frac{24}{We}} \right ). 
\label{alpha_equation1}
\end{equation}
In Eq. (\ref{alpha_equation1}), the radial expansion of a droplet $\alpha=d_{c}/d_{0}$ can be evaluated for a given value of $\We$ at different dimensionless times $\tau$. However, this equation does not consider the falling inertia of the droplet, which can significantly impact the breakup dynamics and the resultant size distribution. Therefore, it is essential to incorporate the effective Weber number ($\We_{\eff}$) in Eq. (\ref{alpha_equation1}). Thus, the modified equation considering the falling inertia of the droplet becomes 
\begin{equation}
\alpha =\textrm{cosh}\left ( \tau \sqrt{\frac{f^{2}}{4}-\frac{24}{\We_{\eff}}} \right ). 
\label{alpha_equation}
\end{equation}
Here, $f$ represents the stretching factor, which is approximately $2\sqrt{2}$ for the cross-flow configuration \cite{kulkarni2014bag}. \ks{The variation of $\alpha$ with $\tau$ is evaluated experimentally using high-speed images. Therefore, in the above equation, the only unknown is $We_{\eff}$, which is determined through curve fitting of the experimental data for different values of $h/D$.} This approach allows us to incorporate the combined effects of size and shape deformation under various experimental conditions. We found that substituting $\We_{\eff} = 12$ in Eq. (\ref{alpha_equation}) yields the best match with the experimental values of $\alpha$ for the retracting-bag breakup mode. Similarly, $\We_{\eff} = 17.5$ aligns well with the experimental values for the bag breakup mode, and $\We_{\eff} = 31$ provides the best agreement for the bag-stamen breakup mode. These values of $\We_{\eff}$ effectively incorporate the combined effects of size and shape deformation under different experimental conditions, allowing the theoretical model to better capture the dynamics observed in the experiments.

\begin{figure}
\centering
\includegraphics[width=0.9\textwidth]{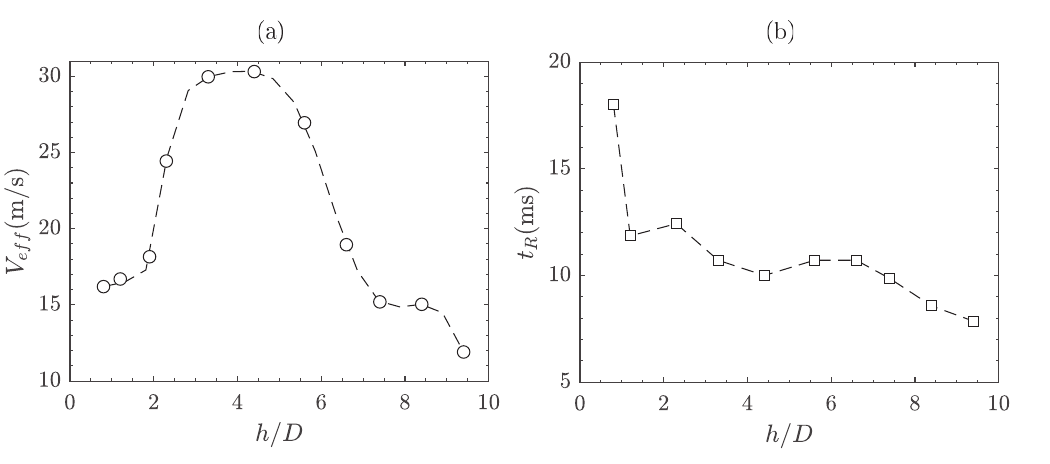}
\caption{Variation of the (a) effective velocity ($V_{\eff}$) and (b) residence time ($t_{R}$) of the droplet with $h/D$. The residence time refers to the duration spent by the droplet in the airstream.}
\label{effective_velocity}
\end{figure}

\begin{figure}
\centering
\includegraphics[width=0.5\textwidth]{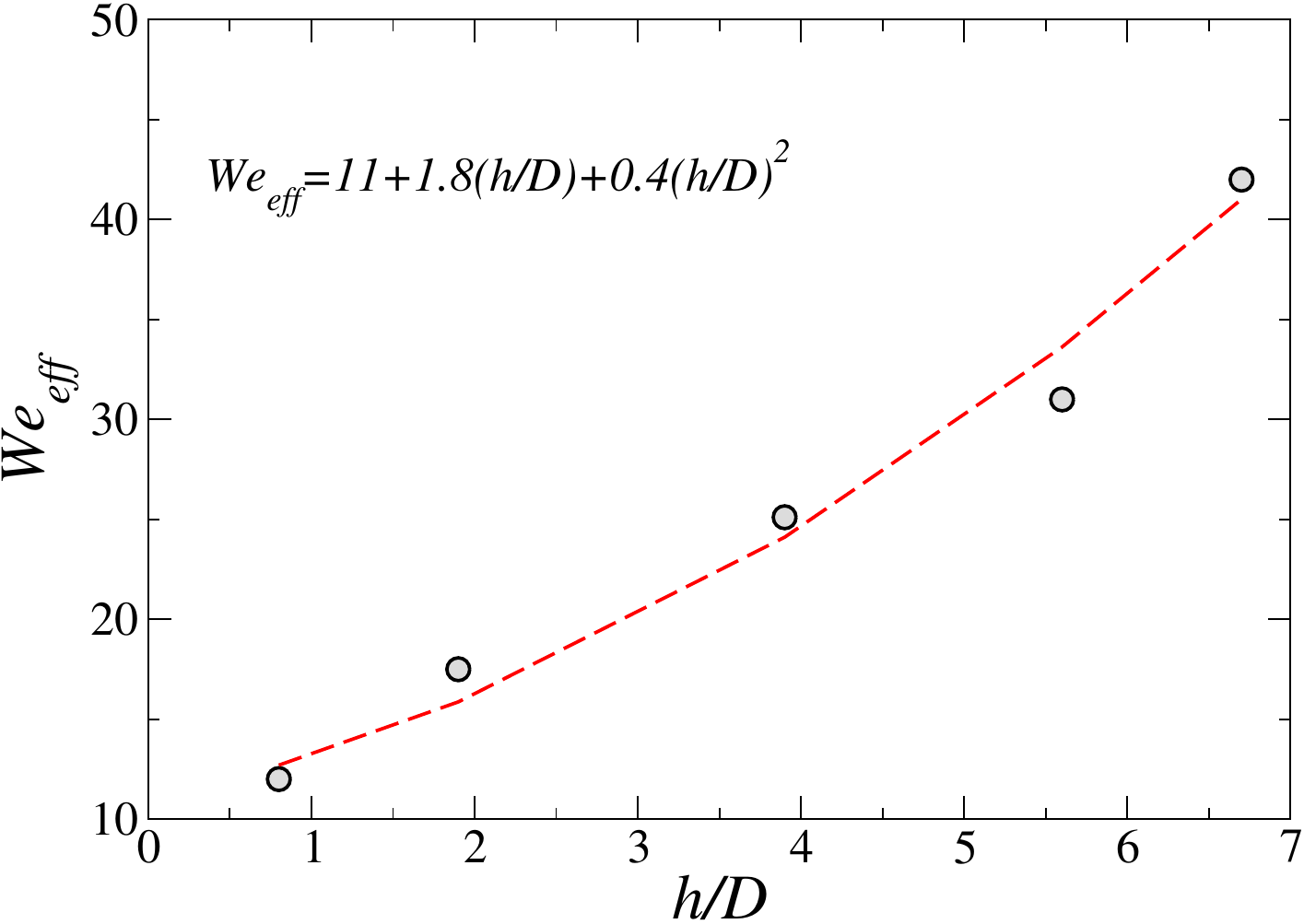} 
\caption{Variation of $\We_{\eff}$ with $h/D$. The dashed line represents $We_{\eff} = 11+1.8 ({h / D})+0.4({h /D})^2$.}
\label{fig6b}
\end{figure}

To gain deeper insights into the analysis of droplet breakup dynamics, we present the results of effective velocity ($V_{\eff}$) and droplet residence time ($t_{R}$) for different values of the normalized height ($h/D$). These results help elucidate the connection between effective velocity and the resident time of the droplet in the airstream. Figure \ref{effective_velocity}(a) illustrates the variation of effective velocity ($V_{\eff}$) with $h/D$. As evident in figure \ref{effective_velocity}(a), $V_{\eff}$ first increases and then decreases. The $V_{\eff}$ represents the relative velocity between the airflow and the droplet \ks{at the point of maximum radial deformation.} The increase in $V_{\eff}$ up to $h/D = 5.6$ is due to higher droplet inertia, causing the drop to experience progressively higher velocity from the outer region to the potential core. Beyond $h/D = 5.6$, droplet inertia increases further, leading to a trend reversal, and $V_{\eff}$ begins to decrease as the drop crosses the potential core, interacting with the lower velocity aerodynamic fields. \ks{It is to be noted that $We_{\eff}$ is not directly evaluated from $V_{\eff}$ and therefore does not exhibit similar trends. This complex interaction of the droplet with the aerodynamic field} leads to different breakup modes. Figure \ref{effective_velocity}(b) illustrates the variation of droplet residence time ($t_{R}$) in milliseconds with $h/D$. The $t_{R}$ represents the time the droplet spends in the aerodynamic field. As shown in figure \ref{effective_velocity}(b), $t_{R}$ decreases with an increase in $h/D$. This decreasing trend is due to the increase in the droplet's falling inertia (and hence the falling velocity) with height. A higher $h/D$ leads to a shorter residence time because the droplet falls faster through the aerodynamic field. Figure \ref{fig6b} illustrates the variation of $\We_{\eff}$ with $h/D$. The best-fit curve, represented as $We_{\eff} = 11+1.8 ({h / D})+0.4({h /D})^2$, qualitatively captures the observed variations for the range of parameters considered in the present study. We utilize the values of the effective Weber number $(\We_{\eff})$ to estimate the droplet size distributions that arise from different breakup modes corresponding to various release heights.

\subsection{Droplet size distribution}

In this section, we present the size distribution of child droplets resulting from the retracting bag, bag, and bag-stamen breakup phenomena. We also compare the experimental results with the analytical model \citep{jackiw2022prediction} by employing the $\We_{\eff}$ instead of $\We$. The volume probability density ($P_v$) is defined as the ratio of the volume occupied by all child droplets of a specific diameter to the total volume occupied by droplets of all diameters. To provide a comprehensive analysis, we incorporate both gamma and log-normal distributions. This dual approach allows for a more robust and detailed comparison with experimental data, thereby providing a deeper understanding of the underlying physical processes.

The volume probability density using the gamma distribution function is given by
\begin{equation} \label{j:eq1}
P_{v}=\frac{\zeta ^{3}P_{n}}{\int_{0}^{\infty}\zeta ^{3}P_{n}d\zeta}={\frac{\zeta ^{3}P_{n}}{\beta^{3}\Gamma (\alpha+3)/\Gamma (\alpha)}},
\end{equation}
where $P_n= {\zeta^{\alpha -1}e^{-\zeta /\beta} / \beta^{\alpha}\Gamma (\alpha)}$; $\zeta \left(=d/d_0\right)$; $\Gamma (\alpha)$ represents the Gamma function; $\alpha=(\bar{\zeta}/\sigma_s)^{2}$ and $\beta=\sigma_s^{2}/\bar{\zeta}$ are the shape and rate parameters, respectively; $\bar{\zeta}$ and $\sigma_s$ are the mean and standard deviation of the distribution, which are estimated based on the characteristic breakup sizes corresponding to each mode.

The volume probability density ($P_v$) for log-normal distribution is given by
\begin{equation} \label{j:aequ1}
P_{v}=\frac{\zeta ^{3}P_{n}}{\int_{0}^{\infty}\zeta ^{3}P_{n}d\zeta}={\frac{\zeta ^{3}P_{n}}{3\mu+4.5\sigma_{l}^2}},
\end{equation}
where, $\zeta \left (=d/d_0 \right)$ and $P_n$ is the number probability density function for log-normal distribution and it can be determined using following equation as 
\begin{equation} \label{j:aequ2}
P_{n}=\frac{1}{\zeta \sigma _{l}\sqrt{2\pi }}\textrm{exp}\left \{ \frac{-(\textrm{log}(\zeta)-\mu )^{2}}{2\sigma _{l}^{2}} \right \}.
\end{equation}
Here, $\mu $ and $\sigma_l$ are the logarithmic mean and logarithmic standard deviation of the distribution, which are estimated based on the characteristic breakup sizes corresponding to each mode.

In the retracting bag, bag and bag-stamen breakup phenomena, the overall size distribution of child droplets is influenced by three distinct modes, namely the node, rim, and bag fragmentation. Hence, it is crucial to account for their respective contributions to the size distribution by performing a weighted summation of each mode. Thus, the total volume probability density ($P_v$) can be determined as
\begin{equation} \label{j:eq3}
P_{v,Total}=w_{N}P_{v,N}+w_{R}P_{v,R}+w_{B}P_{v,B},
\end{equation}
where $w_{N}=V_{N}/V_{0}$, $w_{R}=V_{R}/V_{0}$ and $w_{B}=V_{B}/V_{0}$ represent the contributions of volume weights from the node, rim and bag, respectively. Here, $V_{N}$, $V_{R}$, $V_{B}$ and $V_{0}$ are the node, rim, bag and the initial droplet volumes, respectively. The volume probability density for the node, rim and bag breakup modes are denoted as $P_{v,N}$, $P_{v,R}$ and $P_{v,B}$, respectively. The expressions for the characteristic breakup sizes and the weight contributions for each mode can be found in Appendix \ref{sec:App}. In the following, we present the size distribution of child droplets obtained from three repeated measurements for each set of parameters.

\begin{figure}[h]
\centering
\includegraphics[width=0.9\textwidth]{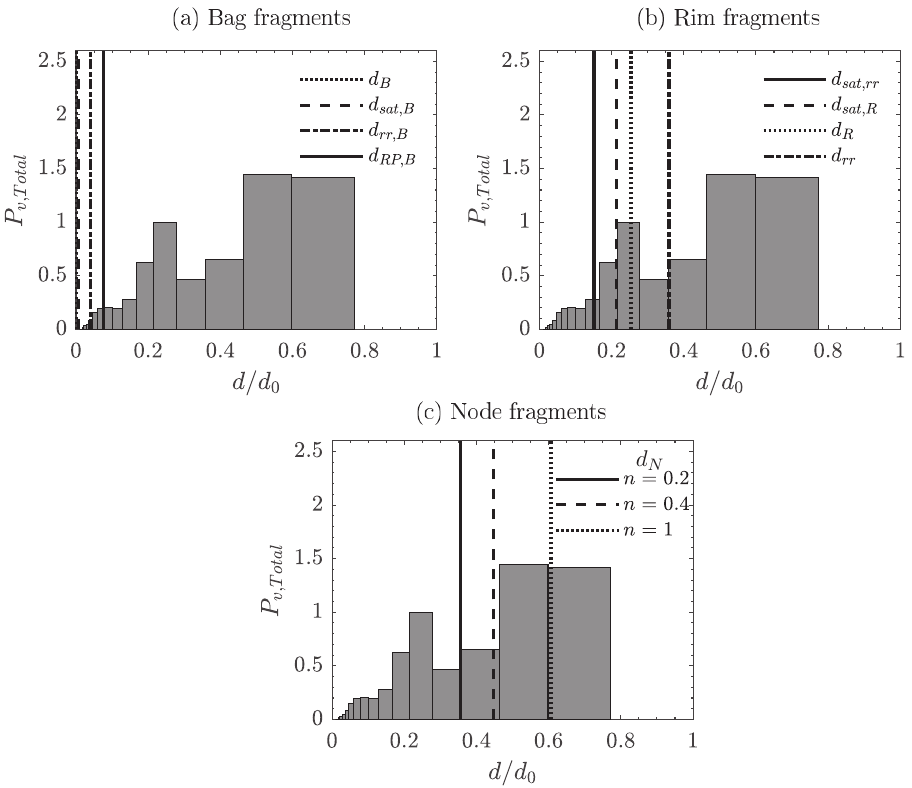}
\caption{Comparison of the theoretically predicted droplet characteristic breakup sizes with experimental data for $h/D = 0.8$ and $\We = 12.1$ (retracting bag breakup with $\We_{\eff} = 12$). Panels (a), (b), and (c) represent the characteristic sizes corresponding to the bag, rim, and node fragmentation processes, respectively. In panel (a), $d_{RP,B}$, $d_{sat,B}$, $d_{B}$, and $d_{rr,B}$ represent characteristic breakup sizes due to Rayleigh-Plateau instability, non-linear instability of liquid ligaments, minimum bag thickness, and receding rim instability, respectively. In panel (b), $d_{R}$ and $d_{rr}$ are the characteristic sizes resulting from capillary and receding rim instabilities, respectively, while $d_{sat,R}$ and $d_{sat,rr}$ are the two breakup sizes due to non-linear instability of liquid ligaments near the pinch-off point. In panel (c), $d_{N}$ represents the characteristic sizes obtained from node breakup, where $n$ is the volume fraction of the node relative to the disk.}
\label{retracting_bag_breakup_sizes}
\end{figure}

Figure \ref{retracting_bag_breakup_sizes}(a-c) illustrates the comparison between theoretically predicted characteristic breakup sizes and experimental data for retracting bag breakup ($h/D = 0.8$ and $\We_{\eff} = 12$). In figure \ref{retracting_bag_breakup_sizes}(a), it can be seen that for the bag fragmentation mode, the rupture of the bag results in four specific breakup sizes: $d_{RP,B}$ due to Rayleigh-Plateau instability, $d_{sat,B}$ due to non-linear instability of liquid ligaments, $d_{B}$ due to minimum bag thickness, and $d_{rr,B}$ due to receding rim instability (see, appendix \S\ref{sec:A3}). These sizes represent the smallest characteristic dimensions governing the analytical size distribution in the bag rupture mode. Figure \ref{retracting_bag_breakup_sizes}(b) shows the characteristic breakup sizes due to rim fragmentation mode. The rim breakup process results in four characteristic sizes: $d_{R}$ due to capillary instability, $d_{rr}$ due to receding rim instability, and $d_{sat,R}$ and $d_{sat,rr}$ due to non-linear instability of liquid ligaments near the pinch-off point (see, appendix \S\ref{sec:A2}). These characteristic sizes are larger than those produced during the bag ruptures and ultimately govern the analytical size distribution in the rim breakup process. Figure \ref{retracting_bag_breakup_sizes}(c) displays the three characteristic sizes, $d_{N}$, corresponding to $n=0.2$, $n=0.4$, and $n=1$, obtained after the node breakup process. The value of $n$ represents the volume fraction of the node relative to the rim (see, appendix \S\ref{sec:A1}). These are the largest characteristic sizes in the entire fragmentation process and govern the size distribution in the node breakup. Thus, figure \ref{retracting_bag_breakup_sizes}(a-c) confirms that the theoretically predicted characteristic sizes align well with the experimental results. Specifically, the peak of the experimental distribution corresponds closely to the mean value of these characteristic sizes.

\begin{figure}[h]
\centering
\includegraphics[width=0.9\textwidth]{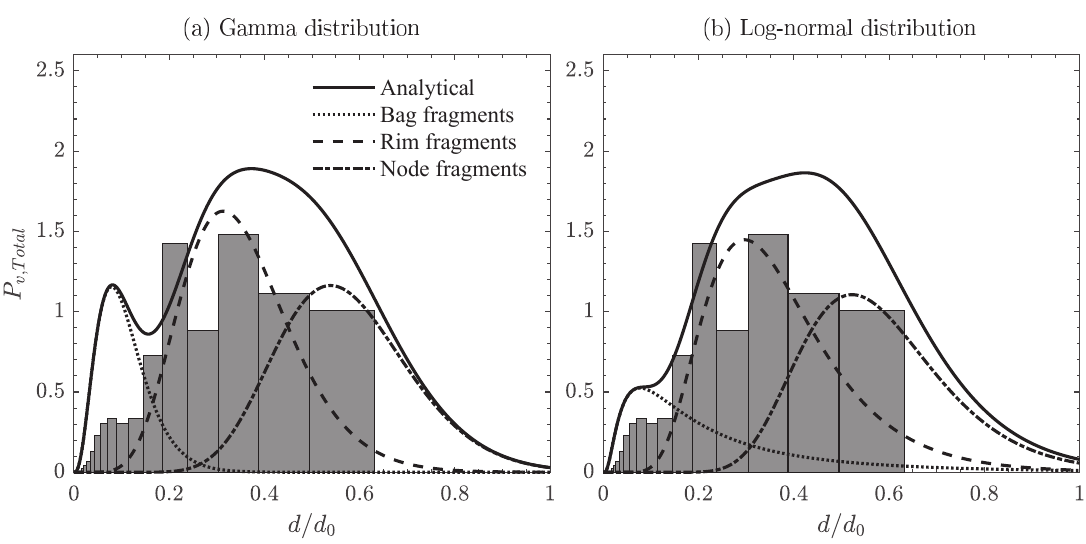}
\caption{Comparison of the volume-weighted density ($P_{v,Total}$) of all fragments at $\tau = 4.78$ (after completion of the breakup phenomenon) obtained from our experiments with the analytical prediction for $h/D=0.8$ and $\We=12.1$. Panels (a) and (b) are associated with the gamma and log-normal distributions, respectively. The analytically predicted individual contributions from different breakup mechanisms, such as bag, rim and node fragmentations, are also depicted. This distribution corresponds to the retracting bag breakup scenario with $\We_{\eff}=12$.}
\label{fig9}
\end{figure}

In figure \ref{fig9}(a), we compare the analytical prediction of the overall size distribution obtained using the gamma distribution function with our experimental results for $h/D=0.8$ at $\tau=4.78$ (an instant after completion of the breakup phenomenon). This case is associated with the retracting bag breakup corresponding to $\We_{\eff}=12$. It can be seen that there are three distinct modes with one marginal and two significant contributions to the size distribution. The solid line indicates the sum of all the modes (bag+rim+nodes) of the size distribution, while the dotted, dashed, and dash-dotted lines represent the contributions from the individual modes. In this breakup scenario, as the droplet stays in the shear layer, the bag fragmentation has a marginal contribution to the overall size distribution. Therefore, the rim and node fragmentation mainly contribute to the overall size distribution. Due to the Rayleigh-Plateau capillary instability, the breakup of the rim generates child droplets with a distribution peak at $d/d_{0} \approx 0.22$, while the node breakup due to Rayleigh-Taylor instability contributes to the droplet size distribution with a peak at $d/d_{0}=0.34$. The theoretical model employing the effective Weber number provides reasonable predictions for rim and node breakup processes. However, it over-predicts the size distribution for the bag breakup process. This is due to the fact that the current model does not account for the retraction of the bag and its subsequent collapse into the rim. Figure \ref{fig9}(b) illustrates the corresponding comparison with the analytical prediction obtained using the log-normal distribution. It is evident that the log-normal distribution accurately predicts the volume contribution of the retracting bag, showing a distribution peak around $d/d_{0} \approx 0.08$. This accuracy is attributed to the long exponential tail characteristic of the log-normal distribution. Furthermore, the log-normal distribution also effectively predicts the volume contributions resulting from rim and node fragmentation.

\begin{figure}[h]
\centering
\includegraphics[width=0.9\textwidth]{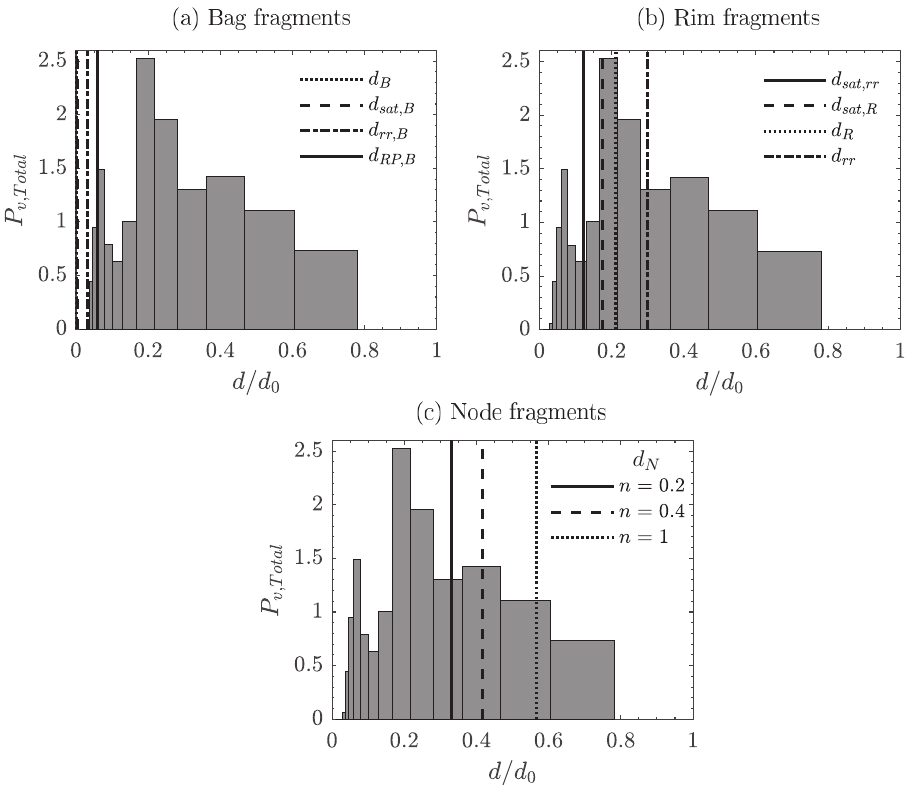}
\caption{Comparison of the theoretically predicted droplet characteristic breakup sizes with experimental data for $h/D = 1.9$ and $\We = 12.1$ (bag breakup with $\We_{\eff} = 17.5$). Panels (a), (b), and (c) represent the characteristic sizes corresponding to the bag, rim, and node fragmentation processes, respectively. In panel (a), $d_{RP,B}$, $d_{sat,B}$, $d_{B}$, and $d_{rr,B}$ represent characteristic breakup sizes due to Rayleigh-Plateau instability, non-linear instability of liquid ligaments, minimum bag thickness, and receding rim instability, respectively. In panel (b), $d_{R}$ and $d_{rr}$ are the characteristic sizes resulting from capillary and receding rim instabilities, respectively, while $d_{sat,R}$ and $d_{sat,rr}$ are the two breakup sizes due to non-linear instability of liquid ligaments near the pinch-off point. In panel (c), $d_{N}$ represents the characteristic sizes obtained from node breakup, where $n$ is the volume fraction of the node relative to the disk.}
\label{bag_breakup_sizes}
\end{figure}

Figure \ref{bag_breakup_sizes}(a-c) depicts the comparison of theoretically predicted characteristic breakup sizes with experimental data for bag breakup ($h/D = 1.9$ and $\We_{\eff} = 17.5$). Figure \ref{bag_breakup_sizes}(a) represents the characteristic sizes of the bag fragmentation mode. In this case, as the droplet interacts with the shear and potential core region, it exhibits the classical bag breakup process. Figure \ref{bag_breakup_sizes}(a) illustrates the four characteristic sizes resulting from Rayleigh-Plateau instability, non-linear instability of liquid ligaments, minimum bag thickness, and receding rim instability in the bag rupture process. These are the smallest characteristic sizes that govern the size distribution for bag fragments. Figure \ref{bag_breakup_sizes}(b) depicts the four characteristic sizes resulting from capillary instability, receding rim instability, and non-linear instability of liquid ligaments near the pinch-off point in the rim breakup process. These moderate characteristic sizes determine the size distribution for rim fragments. Figure \ref{bag_breakup_sizes}(c) shows the three characteristic sizes resulting from Rayleigh-Taylor instability in the node fragmentation process. These are the largest characteristic sizes that govern the size distribution in the node breakup mode.

\begin{figure}[h]
\centering
\includegraphics[width=0.9\textwidth]{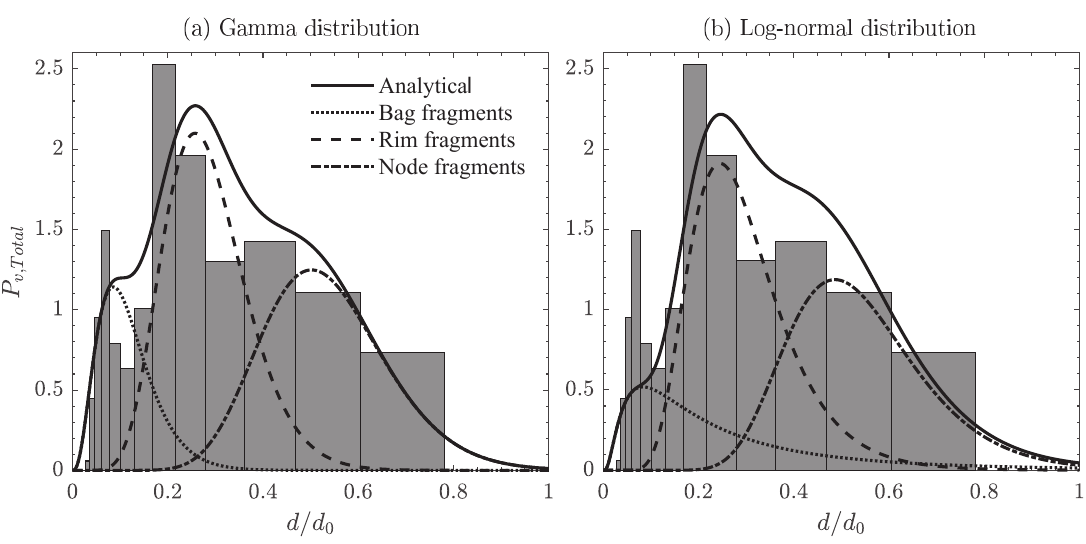}
\caption{Comparison of the volume-weighted density ($P_{v,Total}$) of all fragments at $\tau = 3.37$ (after completion of the breakup phenomenon) obtained from our experiments with the analytical prediction for $h/D=1.9$ and $\We=12.1$. Panels (a) and (b) are associated with the gamma and log-normal distributions, respectively. The analytically predicted individual contributions from different breakup mechanisms, such as bag, rim and node fragmentations, are also depicted. This distribution corresponds to the bag breakup scenario with $\We_{\eff}=17.5$.}
\label{fig10}
\end{figure}

The size distribution of the child droplets at $\tau=3.37$ for $h/D=1.9$ associated with the bag breakup phenomenon for $\We_{\eff}=17.5$ is shown in figure \ref{fig10}(a) and \ref{fig10}(b). Figure \ref{fig10}(a) and \ref{fig10}(b) are associated with the gamma and log-normal distributions, respectively. The theoretical predictions of the size distribution for individual modes and the overall size distribution are also depicted. It can be seen that the bag breakup exhibits three distinct modes in the size distribution of the child droplets. In this case, as the droplet interacts with both the shear and potential core region, the droplet undergoes the normal bag breakup process, which involves the breakup of the inflated bag, rim, and nodes. As a result, all these processes contribute to the overall size distribution. In figure \ref{fig10}(a), the first peak at $d/d_0 \approx 0.07$ is attributed to the bag rupture. The second peak observed at $d/d_0 \approx 0.19$ is due to the fragmentation of the rim, and the third peak at $d/d_0 \approx 0.42$ is due to the node breakup. Inspection of figure \ref{fig10}(a) also reveals that the analytically estimated contributions for bag and rim and node breakups provide reasonable predictions for the experimental size distribution associated with these modes. In addition, the theoretically predicted characteristic sizes are in reasonable agreement with the experimental results. The percentage deviation between the experimental and analytical distributions for bag, rim and node breakup processes are $20\%$, $8.7\%$, and $1.4\%$, respectively. We found that the volume fraction of bag, rim, and nodes are $15\%$, $45\%$, and $40\%$, respectively. Inspection of Figure \ref{fig10}(b) reveals that the log-normal distribution significantly underpredicts the volume contribution of the bag. On the other hand, the analytical distribution reasonably predicts the volume contributions for rim and node fragmentation, with slight discrepancies of $8.5\%$ and $16\%$, respectively. A careful inspection of figure \ref{fig10}(a) and \ref{fig10}(b) also reveals that the gamma distribution accurately predicts the volume contribution for smaller child droplets due to its short exponential tail, whereas the log-normal distribution predicts the volume contribution for larger child droplets reasonably well due to its long exponential tail.

\begin{figure}[h]
\centering
\includegraphics[width=0.9\textwidth]{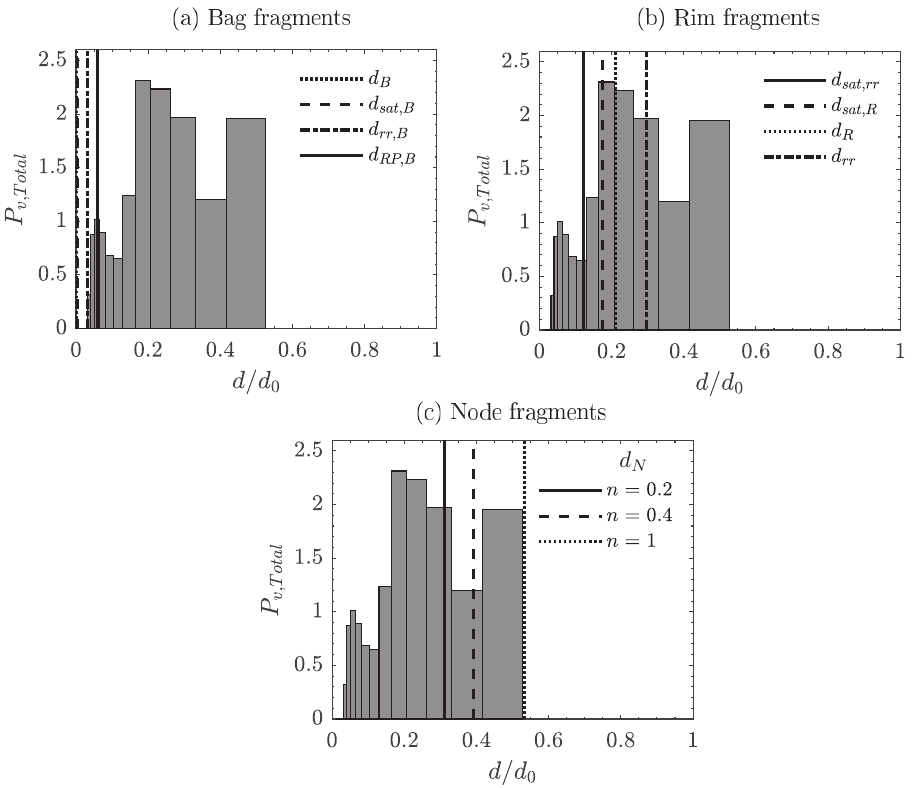}
\caption{Comparison of the theoretically predicted droplet characteristic breakup sizes with experimental data for $h/D = 5.6$ and $We = 12.1$ (bag-stamen breakup with $\We_{\eff} = 31$). Panels (a), (b), and (c) represent the characteristic sizes corresponding to the bag, rim, and node fragmentation processes, respectively. In panel (a), $d_{RP,B}$, $d_{sat,B}$, $d_{B}$, and $d_{rr,B}$ represent characteristic breakup sizes due to Rayleigh-Plateau instability, non-linear instability of liquid ligaments, minimum bag thickness, and receding rim instability, respectively. In panel (b), $d_{R}$ and $d_{rr}$ are the characteristic sizes resulting from capillary and receding rim instabilities, respectively, while $d_{sat,R}$ and $d_{sat,rr}$ are the two breakup sizes due to non-linear instability of liquid ligaments near the pinch-off point. In panel (c), $d_{N}$ represents the characteristic sizes obtained from node breakup, where $n$ is the volume fraction of the node relative to the disk.}
\label{bag_stamen_sizes}
\end{figure}

Figure \ref{bag_stamen_sizes}(a-c) compares theoretically predicted characteristic breakup sizes with experimental data for bag-stamen breakup ($h/D = 5.6$ and $\We_{\eff} = 31$). In this case, the droplet interacts with the potential core of the flow field. Figure \ref{bag_stamen_sizes}(a) represents the four characteristic sizes resulting from Rayleigh-Plateau instability, non-linear instability of liquid ligaments, minimum bag thickness, and receding rim instability during the bag rupture process. These sizes are the smallest and dictate the size distribution for bag fragments. Figure \ref{bag_stamen_sizes}(b) shows the four characteristic sizes resulting from capillary instability, receding rim instability, and non-linear instability of liquid ligaments near the pinch-off point in the rim breakup process. These sizes are moderate and determine the size distribution for rim fragments. Figure \ref{bag_stamen_sizes}(c) presents the three characteristic sizes resulting from Rayleigh-Taylor instability in the node fragmentation process. These are the largest sizes of child droplets and mainly contribute to the size distribution of the node breakup mode.

\begin{figure}
\centering
\includegraphics[width=0.9\textwidth]{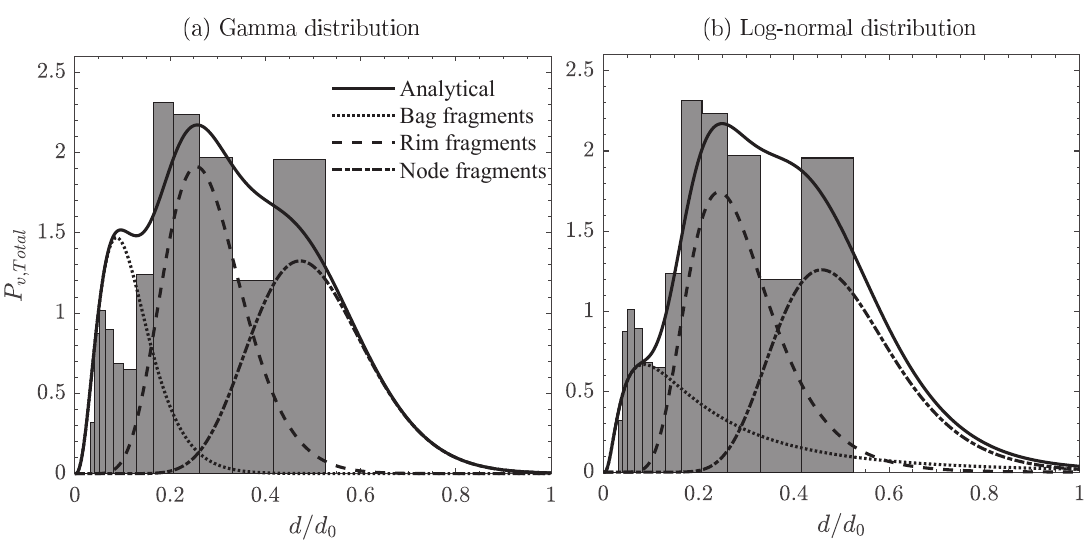}
\caption{Comparison of the volume-weighted density ($P_{v,Total}$) of all fragments at $\tau = 3.02$ (after completion of the breakup phenomenon) obtained from our experiments with the analytical prediction for $h/D=5.6$ and $\We=12.1$. Panels (a) and (b) are associated with the gamma and log-normal distributions, respectively. The analytically predicted individual contributions from different breakup mechanisms, such as bag, rim and node fragmentations, are also depicted. This distribution corresponds to the bag-stamen breakup scenario with $\We_{\eff}=31$.}
\label{fig11}
\end{figure}

Figure \ref{fig11}(a) shows the analytical prediction of the overall distribution obtained using the gamma distribution function with the experimental results for $h/D=5.6$ associated with the bag-stamen breakup at $\We_{\eff}=31$. Similar to the bag breakup ($h/D=1.9$), the bag-stamen breakup ($h/D=5.6$) undergoes three different modes in the size distribution. For $h/D=5.6$, the droplet completely migrates into the potential core region of the airstream and undergoes the bag-stamen breakup process, which involves the breakup of the inflated bag, rim stamen, and nodes. As a result, all these processes contribute to the overall size distribution. In figure \ref{fig11}(a), the first peak at $d/d_0 \approx 0.058$ corresponds to the fragmentation of the bag. The second peak observed at $d/d_0 \approx 0.18$ is due to the fragmentation of the rim, and the third peak at $d/d_0 \approx 0.47$ is due to the node breakup. The comparison with the analytical prediction reveals that the contributions of the bag and node to the volume-weight size distributions are over-estimated and under-estimated, respectively. On the other hand, the contribution of rim fragmentation is slightly under-predicted.  This discrepancy arises because the model does not account for the fragmentation of the stamen, which is similar in size to the rim. Furthermore, since the volume of the bag contributes to the formation of the stamen, the model also falls short in predicting the extent of child droplets produced by the bag breakup. However, the theoretically predicted characteristic sizes reasonably agree with the experimental results. We found that the volume fraction of bag, rim, and nodes are $20\%$, $40\%$ and $40\%$, respectively, for $h/D=5.6$. In figure \ref{fig11}(b), we observe that log-normal distribution underestimates the volume contribution of the bag by $42.8\%$, but the analytical distribution accurately predicts the volume contributions for rim and node fragmentation, with minor discrepancies of $4.5\%$ and $18\%$, respectively. It can be seen in figure \ref{fig11}(a) and (b) that the gamma distribution accurately predicts the volume contribution for smaller child droplets due to its short exponential tail, while the log-normal distribution reasonably predicts the volume contribution for larger child droplets due to its long exponential tail.

In summary, while the model initially developed by \citet{jackiw2022prediction} for bag breakup in a cross-flow configuration involving droplets released with negligible inertia into the airstream, the present study enhances the analytical model by incorporating the effective Weber number. This addition allows the model to account for the inertia effects of droplets released from varying heights, enabling it to predict size distributions for a range of breakup phenomena, including retracting bag, bag and bag-stamen breakups. The size distribution of the retracting bag is primarily influenced by rim and node fragmentation, resulting in a bimodal distribution. In contrast, the bag and bag-stamen breakups yield a tri-modal size distribution due to the collective contributions of the three breakup modes: bag, rim, and node.

\clearpage

\section{Concluding remarks}\label{sec:conc}

We investigate the effect of the inertia of a droplet, caused by its free fall from various heights, on its morphology, breakup and the resulting size distribution of the child droplets when it encounters an airstream using high-speed imaging and in-line holography techniques. In contrast to previous studies that considered neglecting the inertia of a spherical droplet when introduced into the potential core region of the airstream, a droplet dispensed from a height undergoes an accelerating phase during its free fall. Thus, a complex interplay between its weight, the aerodynamic forces of the airstream, and the inertia of the droplet influences its subsequent dynamics. 

Our findings reveal that a droplet falling from varying heights displays distinct breakup patterns, even when exposed to a constant airstream (maintaining $\We=12.1$). It is well known that a droplet undergoes a transition from the vibrational to the bag breakup at $\We \approx 12$ when introduced in the potential core region of the airstream. For the parameters considered in the present study, we found that the droplet experiences vibrational breakup when introduced at $h/D=0.6$ (at a location slightly above the air nozzle). In this scenario, due to its low inertia, the droplet cannot penetrate the airstream and exhibits the vibrational mode near the outer periphery of the airflow, which is caused by a weak aerodynamic force. In contrast, increasing the release height of the droplet leads to a sequence of breakup modes, transitioning from vibrational breakup to retracting bag breakup ($h/D=0.8$), bag breakup ($h/D=1.9$), bag-stamen breakup ($h/D=5.6$), retracting bag-stamen breakup ($h/D=8.4$), and eventually returning to vibrational breakup ($h/D=11.2$). As the release height increases, the inertia of the droplet also increases, causing it to migrate into the potential core region. For $h/D=11.2$, the droplet possesses enough momentum to traverse the potential core region. Consequently, the droplet displays vibrational breakup in the shear zone located in the lower part of the airstream. We also found that the droplet undergoes natural oscillations in its shape, transitioning between oblate, spherical, and prolate shapes as it descends from a height. These shape oscillations arise from the interplay between inertia and surface tension forces, resembling the fundamental mode of the theoretically predicted oscillations observed in a freely suspended droplet \citep{rayleigh1879capillary}. 

We observed that the radial deformation of the droplet when encountering the same airstream (fixed Weber number defined based on the air velocity at the nozzle exit) undergoes significant alterations when released from different heights. This deformation plays a crucial role in the breakup morphology and in determining the sizes of the resulting child droplets. Consequently, we have incorporated an effective Weber number ($\We_{\eff}$) that accounts for the radial deformation in the analytical model \citep{jackiw2022prediction} to estimate the droplet size distribution for different release heights. This approach adequately predicts the size distribution of child droplets for various breakup modes observed at different heights. Our findings also highlight that the size distribution resulting from the retracting bag breakup primarily results from rim and node fragmentation, leading to a bimodal distribution. In contrast, bag and bag-stamen breakups yield a tri-modal size distribution due to the combined contributions of three breakup modes: bag, rim, and node.

To the best of our knowledge, the influence of droplet inertia when interacting with an airstream, a factor relevant in numerous practical applications and natural phenomena, has hitherto remained largely unexplored. Therefore, the present study, which investigates the distinct morphologies, breakup phenomena and the resulting size distribution of child droplets as the primary droplet descends freely from a height and encounters an airstream, offers a novel perspective and holds significance in enhancing our understanding of numerous practical applications and natural phenomena.\\

\vspace{4mm}
\noindent{\bf Declaration of Interests:} The authors report no conflict of interest. \\
\\
\noindent{\bf Acknowledgement:} {L.D.C. and K.C.S. thank the Science \& Engineering Research Board, India, for their financial support through grants SRG/2021/001048 and CRG/2020/000507, respectively. We also thank IIT Hyderabad for the financial support through grant IITH/CHE/F011/SOCH1. S.S.A. also thanks the PMRF Fellowship.}

\appendix
\section{Analytical model for droplet size distribution} 
\label{sec:App}

In this section, we provide the volume weights and characteristic sizes associated with each mode, specifically bag, rim, and node, as outlined in \citet{jackiw2021aerodynamic} and \citet{jackiw2022prediction}. 

The node breakup volume weight, $w_N$, can be calculated as follows: 
\begin{equation} \label{j:eq4}
w_{N}=\frac{V_{N}}{V_{0}}=\frac{V_{D}}{V_{0}}\frac{V_{N}}{V_{D}},
\end{equation}
where the disk volume $V_D$ is given by \citep{jackiw2021aerodynamic}
\begin{equation} \label{j:eq5a}
\frac{V_{D}}{V_{0}}=\frac{3}{2}\left [ \left ( \frac{2R_{i}}{d_{0}} \right )^{2}\left ( \frac{h_{i}}{d_{0}} \right )-2\left ( 1-\frac{\pi }{4} \right )\left ( \frac{2R_{i}}{d_{0}} \right )\left ( \frac{h_{i}}{d_{0}} \right )^{2} \right ].
\end{equation}

In Eq. (\ref{j:eq5a}), $h_i$ denotes the thickness of the disk, and $2R_i$ represents the major diameter of the rim. The expression for calculation of  $h_i$ and $2R_i$ are given as follows: \citep{jackiw2021aerodynamic,jackiw2022prediction}
\begin{equation} \label{j:eq6}
\frac{h_{i}}{d_{0}}=\frac{4}{\We_{rim}+10.4},
\end{equation}
and 
\begin{equation} \label{j:eq7}
\frac{2R_{i}}{d_{0}}=1.63-2.88e^{({-0.312\We_{\eff}})}.
\end{equation}

In Eqs. (\ref{j:eq6}) and (\ref{j:eq7}), $\We_{rim}$ represents the rim Weber number, which signifies the balance between the radial momentum generated at the outer edge of the droplet and the surface tension that stabilizes the droplet. This parameter is calculated using the formula $\We_{rim}=\rho_{w} \dot{R}^{2}{d_{0}}/\sigma$. The constant radial expansion rate of a droplet ($\dot{R}$) can be determined as 
$\dot{R} = \frac{1.125}{2} \left(\frac{U \sqrt{\rho_a/\rho_w}}{1-32/9 \We_{\eff}}\right)$ \citep{jackiw2022prediction}.
The ratio of $V_{N}/V_{D}$ indicates the volume fraction between the nodes and the disk, which is approximately equal to $0.4$ according to \citet{jackiw2022prediction}.

The volume weight of the rim, $w_R$, can be calculated from the following equation \citep{jackiw2021aerodynamic}:
\begin{equation} \label{j:eq5}
w_{R}=\frac{V_{R}}{V_{0}}=\frac{3\pi }{2}\left [ \left ( \frac{2R_{i}}{d_{0}} \right )\left ( \frac{h_{i}}{d_{0}} \right )^{2}-\left ( \frac{h_{i}}{d_{0}} \right )^{3} \right ],
\end{equation}

The volume weight of the bag, $w_B$ is given by:
\begin{equation} \label{j:eq8}
w_B=\frac{V_{B}}{V_{0}}=\frac{V_{D}}{V_{0}}-\frac{V_{N}}{V_{0}}-\frac{V_{R}}{V_{0}}.
\end{equation}

The aforementioned discussion describes the method for the calculation of volume weights for each mode. To obtain both individual and overall size distributions, it is essential to determine the characteristic droplet sizes for each mode. Therefore, in the subsequent sections, we present the estimation of these characteristic sizes for each mode.

\subsection{Node droplet sizes ($d_N$)}\label{sec:A1}
The nodes are generated on the rim due to the Rayleigh-Taylor (RT) instability, where the lighter fluid (air phase) displaces the heavier fluid (liquid phase) \citep{zhao2010morphological}. The droplet size ($d_N$) resulting from node breakup, as per the RT instability theory, is described by \citep{jackiw2022prediction} as follows:
\begin{equation} \label{j:eq9}
\frac{d_{N}}{d_{0}}=\left [ \frac{3}{2}\left ( \frac{h_{i}}{d_{0}} \right )^{2}\frac{\lambda_{RT} }{d_{0}}n \right ]^{1/3},
\end{equation}
where, $n=V_{N}/V_{D}$ signifies the volume fraction of the nodes compared to the disk. \citet{jackiw2022prediction} approximated that the minimum, mean, and maximum values for $n$ are 0.2, 0.4, and 1, respectively. Using these three values of $n$,  the three characteristic sizes of the node droplets can be calculated. Subsequently, the number-based mean and standard deviation for node breakup can be determined based on these three characteristic sizes. The maximum susceptible wavelength of the RT instability is given as, $\lambda_{RT}=2\pi\sqrt{3\sigma/\rho_{w} a}$, such that $a=\frac{3}{4}C_{D}\frac{U^{2}}{d_{0}}\frac{\rho_{a} }{\rho_{w}}\left ({D_{max}/d_{0}} \right )^{2}$ is the acceleration of the deforming droplet. According to \citet{zhao2010morphological}, the drag coefficient ($C_{D}$) of the disk shape droplet is about 1.2, and the extent of droplet deformation is expressed as ${D_{max} / d_0}={2 / (1+\exp{(-0.0019 {\We_{\eff}}^{2.7})})}$.
 
\subsection{Rim droplet sizes ($d_R$)} \label{sec:A2}
The child droplets formed during the rim breakup result from three main mechanisms: the Rayleigh-Plateau instability, the receding rim instability as described by \citet{jackiw2022prediction}, and the nonlinear instability of liquid ligaments near the pinch-off point.

The size of child droplets ($d_R$) resulting from the Rayleigh-Plateau instability mechanism is as follows:
\begin{equation} \label{j:eq14}
\frac{d_{R}}{d_{0}}=1.89\frac{h_{f}}{d_{0}}.
\end{equation} 
Here, $h_{f}$ is the final rim thickness, which is given by
\begin{equation} \label{j:eq15}
\frac{h_{f}}{d_{0}}=\frac{h_{i}}{d_{0}}\sqrt{\frac{R_{i}}{R_f}},
\end{equation}
where $R_f$ is the bag radius at the time of its burst and it can be evaluated as \citep{kirar2022experimental}
\begin{equation} \label{j:eq16}
R_f=\frac{d_{0}}{2\eta} \left [ 2e^{\tau ^{\prime}\sqrt{p}}+\left ( \frac{\sqrt{p}}{\sqrt{q}}-1 \right )e^{-\tau ^{\prime}\sqrt{q}}-\left ( \frac{\sqrt{p}}{\sqrt{q}}+1 \right )e^{\tau ^{\prime}\sqrt{q}} \right ],
\end{equation}   
where, $\eta = f^{2}-120/\We_{\eff}$, $p=f^{2}-96/\We_{\eff}$ and $q=24/\We_{\eff}$. 
In equation (\ref{j:eq16}), the dimensionless time, denoted as $\tau^{\prime}$, is defined as the ratio of the bursting time ($t_b$) to the characteristic deformation time ($t_d$). These times can obtained as follows \citep{jackiw2022prediction}:
\begin{equation} \label{j:eq17}
t_{b}=\frac{\left [ \left ( \frac{2R_{i}}{d_{0}} \right )-2\left ( \frac{h_{i}}{d_{0}} \right ) \right ]}{\frac{2\dot{R}}{d_{0}}}\left [ -1+\sqrt{1+9.4\frac{8t_{d}}{\sqrt{3\We_{\eff}}}\frac{\frac{2\dot{R}}{d_{0}}}{\left [ \left ( \frac{2R_{i}}{d_{0}} \right )-2\left ( \frac{h_{i}}{d_{0}} \right ) \right ]}\sqrt{\frac{V_{B}}{V_{0}}}} \right ],
\end{equation} 
and 
\begin{equation} \label{j:eq18}
t_{d}=\frac{d_0}{U}\sqrt{\frac{\rho_{w} }{\rho _{a}}}.
\end{equation}

The second mechanism contributing to rim breakup is the receding rim instability, as described by \citet{jackiw2022prediction}. This mechanism yields a droplet size ($d_{rr}$) determined as follows:
\begin{equation} \label{j:eq19}
\frac{d_{rr}}{d_{0}}=\left [ \frac{3}{2}\left ( \frac{h_{f}}{d_{0}} \right )^{2}\frac{\lambda _{rr}}{d_{0}} \right ]^{1/3}.
\end{equation}
Here, $\lambda_{rr}$ represents the wavelength of the receding rim instability and is calculated as $\lambda_{rr} = 4.5b_{rr}$. Here, $b_{rr}$ signifies the thickness of the receding rim and is determined by $b_{rr} = \sqrt{\sigma / (\rho_{w} a_{rr})}$, with $a_{rr} = U_{rr}^{2}/R_{f}$ representing the acceleration of the receding rim as described by \citet{wang2018universal}. The receding rim velocity, denoted as $U_{rr}$, is determined experimentally.

The rim breakup is a consequence of the nonlinear instability of liquid ligaments near the pinch-off point. To determine the characteristic size associated with this mechanism, it is necessary to consider both the Rayleigh-Plateau and receding rim instabilities, which are expressed as follows\citep{keshavarz2020rotary}
\begin{equation} \label{j:eq20}
d_{sat,R}=\frac{d_R}{\sqrt{2+3Oh_{R}/\sqrt{2}}} ~~{\rm and}
\end{equation}   
\begin{equation} \label{j:eq21}
d_{sat,{rr}}=\frac{d_{rr}}{\sqrt{2+3Oh_{R}/\sqrt{2}}},
\end{equation}
respectively. 
In this context, $Oh_{R}$ is defined as the Ohnesorge number based on the final rim thickness, calculated as $Oh_{R} = \mu / \sqrt{\rho_{w} h_{f}^{3}\sigma}$. The characteristic sizes described in Equations (\ref{j:eq14}), (\ref{j:eq19}), (\ref{j:eq20}), and (\ref{j:eq21}) are employed to determine the number-based mean and standard deviation for the rim breakup.

\subsection{Bag droplet sizes ($d_{B}$)} \label{sec:A3}
The factors influencing the droplet size distribution resulting from the rupture of a bag film include the minimum bag thickness, the receding rim thickness ($b_{rr}$), the Rayleigh-Plateau instability, and the nonlinear instability of liquid ligaments. These factors collectively give rise to four characteristic sizes for the satellite droplets, which are expressed as \cite{jackiw2022prediction},
\begin{equation} \label{j:eq21_new}
 d_{B}=h_{min}, 
   \end{equation}  
   \begin{equation} \label{j:eq22}
 d_{rr,B}=b_{rr},
   \end{equation}  
    \begin{equation} \label{j:eq23}
 d_{RP,B}=1.89 b_{rr},
   \end{equation}  
   \begin{equation} \label{j:eq24}
d_{sat,B}=\frac{d_{RP,B}}{\sqrt{2+3Oh_{rr}/\sqrt{2}}}.
   \end{equation} 
\citet{jackiw2022prediction} found that $h_{min} = \pm 2.3$ $\mu \textrm{m}$. In the above equations, $Oh_{rr}$ represents the Ohnesorge number based on the receding rim thickness, denoted as $b_{rr}$. These characteristic sizes are employed to calculate the number-based mean and standard deviation associated with the bag fragmentation mode.


\begin{thebibliography}{44}%
\makeatletter
\providecommand \@ifxundefined [1]{%
 \@ifx{#1\undefined}
}%
\providecommand \@ifnum [1]{%
 \ifnum #1\expandafter \@firstoftwo
 \else \expandafter \@secondoftwo
 \fi
}%
\providecommand \@ifx [1]{%
 \ifx #1\expandafter \@firstoftwo
 \else \expandafter \@secondoftwo
 \fi
}%
\providecommand \natexlab [1]{#1}%
\providecommand \enquote  [1]{``#1''}%
\providecommand \bibnamefont  [1]{#1}%
\providecommand \bibfnamefont [1]{#1}%
\providecommand \citenamefont [1]{#1}%
\providecommand \href@noop [0]{\@secondoftwo}%
\providecommand \href [0]{\begingroup \@sanitize@url \@href}%
\providecommand \@href[1]{\@@startlink{#1}\@@href}%
\providecommand \@@href[1]{\endgroup#1\@@endlink}%
\providecommand \@sanitize@url [0]{\catcode `\\12\catcode `\$12\catcode
  `\&12\catcode `\#12\catcode `\^12\catcode `\_12\catcode `\%12\relax}%
\providecommand \@@startlink[1]{}%
\providecommand \@@endlink[0]{}%
\providecommand \url  [0]{\begingroup\@sanitize@url \@url }%
\providecommand \@url [1]{\endgroup\@href {#1}{\urlprefix }}%
\providecommand \urlprefix  [0]{URL }%
\providecommand \Eprint [0]{\href }%
\providecommand \doibase [0]{https://doi.org/}%
\providecommand \selectlanguage [0]{\@gobble}%
\providecommand \bibinfo  [0]{\@secondoftwo}%
\providecommand \bibfield  [0]{\@secondoftwo}%
\providecommand \translation [1]{[#1]}%
\providecommand \BibitemOpen [0]{}%
\providecommand \bibitemStop [0]{}%
\providecommand \bibitemNoStop [0]{.\EOS\space}%
\providecommand \EOS [0]{\spacefactor3000\relax}%
\providecommand \BibitemShut  [1]{\csname bibitem#1\endcsname}%
\let\auto@bib@innerbib\@empty
\bibitem [{\citenamefont {Villermaux}(2007)}]{villermaux2007fragmentation}%
  \BibitemOpen
  \bibfield  {author} {\bibinfo {author} {\bibfnamefont {E.}~\bibnamefont
  {Villermaux}},\ }\bibfield  {title} {\bibinfo {title} {Fragmentation},\
  }\href@noop {} {\bibfield  {journal} {\bibinfo  {journal} {Annu. Rev. Fluid
  Mech.}\ }\textbf {\bibinfo {volume} {39}},\ \bibinfo {pages} {419} (\bibinfo
  {year} {2007})}\BibitemShut {NoStop}%
\bibitem [{\citenamefont {Jain}\ \emph {et~al.}(2019)\citenamefont {Jain},
  \citenamefont {Tyagi}, \citenamefont {Prakash}, \citenamefont {Ravikrishna},\
  and\ \citenamefont {Tomar}}]{jain2019secondary}%
  \BibitemOpen
  \bibfield  {author} {\bibinfo {author} {\bibfnamefont {S.~S.}\ \bibnamefont
  {Jain}}, \bibinfo {author} {\bibfnamefont {N.}~\bibnamefont {Tyagi}},
  \bibinfo {author} {\bibfnamefont {R.~S.}\ \bibnamefont {Prakash}}, \bibinfo
  {author} {\bibfnamefont {R.~V.}\ \bibnamefont {Ravikrishna}},\ and\ \bibinfo
  {author} {\bibfnamefont {G.}~\bibnamefont {Tomar}},\ }\bibfield  {title}
  {\bibinfo {title} {Secondary breakup of drops at moderate weber numbers:
  Effect of density ratio and {R}eynolds number},\ }\href@noop {} {\bibfield
  {journal} {\bibinfo  {journal} {Int. J. Multiphase Flow}\ }\textbf {\bibinfo
  {volume} {117}},\ \bibinfo {pages} {25} (\bibinfo {year} {2019})}\BibitemShut
  {NoStop}%
\bibitem [{\citenamefont {Villermaux}(2020)}]{villermaux2020fragmentation}%
  \BibitemOpen
  \bibfield  {author} {\bibinfo {author} {\bibfnamefont {E.}~\bibnamefont
  {Villermaux}},\ }\bibfield  {title} {\bibinfo {title} {Fragmentation versus
  cohesion},\ }\href@noop {} {\bibfield  {journal} {\bibinfo  {journal} {J.
  Fluid Mech.}\ }\textbf {\bibinfo {volume} {898}} (\bibinfo {year}
  {2020})}\BibitemShut {NoStop}%
\bibitem [{\citenamefont {Hopfes}\ \emph {et~al.}(2021)\citenamefont {Hopfes},
  \citenamefont {Petersen}, \citenamefont {Wang}, \citenamefont {Giglmaier},\
  and\ \citenamefont {Adams}}]{hopfes2021secondary}%
  \BibitemOpen
  \bibfield  {author} {\bibinfo {author} {\bibfnamefont {T.}~\bibnamefont
  {Hopfes}}, \bibinfo {author} {\bibfnamefont {J.}~\bibnamefont {Petersen}},
  \bibinfo {author} {\bibfnamefont {Z.}~\bibnamefont {Wang}}, \bibinfo {author}
  {\bibfnamefont {M.}~\bibnamefont {Giglmaier}},\ and\ \bibinfo {author}
  {\bibfnamefont {N.~A.}\ \bibnamefont {Adams}},\ }\bibfield  {title} {\bibinfo
  {title} {Secondary atomization of liquid metal droplets at moderate weber
  numbers},\ }\href@noop {} {\bibfield  {journal} {\bibinfo  {journal} {Int. J.
  Multiphase Flow}\ }\textbf {\bibinfo {volume} {143}},\ \bibinfo {pages}
  {103723} (\bibinfo {year} {2021})}\BibitemShut {NoStop}%
\bibitem [{\citenamefont {Raut}\ \emph {et~al.}(2021)\citenamefont {Raut},
  \citenamefont {Konwar}, \citenamefont {Murugavel}, \citenamefont {Kadge},
  \citenamefont {Gurnule}, \citenamefont {Sayyed}, \citenamefont {Todekar},
  \citenamefont {Malap}, \citenamefont {Bankar},\ and\ \citenamefont
  {Prabhakaran}}]{raut2021microphysical}%
  \BibitemOpen
  \bibfield  {author} {\bibinfo {author} {\bibfnamefont {B.~A.}\ \bibnamefont
  {Raut}}, \bibinfo {author} {\bibfnamefont {M.}~\bibnamefont {Konwar}},
  \bibinfo {author} {\bibfnamefont {P.}~\bibnamefont {Murugavel}}, \bibinfo
  {author} {\bibfnamefont {D.}~\bibnamefont {Kadge}}, \bibinfo {author}
  {\bibfnamefont {D.}~\bibnamefont {Gurnule}}, \bibinfo {author} {\bibfnamefont
  {I.}~\bibnamefont {Sayyed}}, \bibinfo {author} {\bibfnamefont
  {K.}~\bibnamefont {Todekar}}, \bibinfo {author} {\bibfnamefont
  {N.}~\bibnamefont {Malap}}, \bibinfo {author} {\bibfnamefont
  {S.}~\bibnamefont {Bankar}},\ and\ \bibinfo {author} {\bibfnamefont
  {T.}~\bibnamefont {Prabhakaran}},\ }\bibfield  {title} {\bibinfo {title}
  {Microphysical origin of raindrop size distributions during the indian
  monsoon},\ }\href@noop {} {\bibfield  {journal} {\bibinfo  {journal}
  {Geophys. Res. Lett.}\ }\textbf {\bibinfo {volume} {48}},\ \bibinfo {pages}
  {e2021GL093581} (\bibinfo {year} {2021})}\BibitemShut {NoStop}%
\bibitem [{\citenamefont {Xu}\ \emph {et~al.}(2022)\citenamefont {Xu},
  \citenamefont {Wang},\ and\ \citenamefont {Che}}]{xu2022droplet}%
  \BibitemOpen
  \bibfield  {author} {\bibinfo {author} {\bibfnamefont {Z.}~\bibnamefont
  {Xu}}, \bibinfo {author} {\bibfnamefont {T.}~\bibnamefont {Wang}},\ and\
  \bibinfo {author} {\bibfnamefont {Z.}~\bibnamefont {Che}},\ }\bibfield
  {title} {\bibinfo {title} {Droplet breakup in airflow with strong shear
  effect},\ }\href@noop {} {\bibfield  {journal} {\bibinfo  {journal} {J. Fluid
  Mech.}\ }\textbf {\bibinfo {volume} {941}},\ \bibinfo {pages} {A54} (\bibinfo
  {year} {2022})}\BibitemShut {NoStop}%
\bibitem [{\citenamefont {Kant}\ \emph {et~al.}(2023)\citenamefont {Kant},
  \citenamefont {Pairetti}, \citenamefont {Saade}, \citenamefont {Popinet},
  \citenamefont {Zaleski},\ and\ \citenamefont {Lohse}}]{kant2022bags}%
  \BibitemOpen
  \bibfield  {author} {\bibinfo {author} {\bibfnamefont {P.}~\bibnamefont
  {Kant}}, \bibinfo {author} {\bibfnamefont {C.}~\bibnamefont {Pairetti}},
  \bibinfo {author} {\bibfnamefont {Y.}~\bibnamefont {Saade}}, \bibinfo
  {author} {\bibfnamefont {S.}~\bibnamefont {Popinet}}, \bibinfo {author}
  {\bibfnamefont {S.}~\bibnamefont {Zaleski}},\ and\ \bibinfo {author}
  {\bibfnamefont {D.}~\bibnamefont {Lohse}},\ }\bibfield  {title} {\bibinfo
  {title} {Bag-mediated film atomization in a cough machine},\ }\href@noop {}
  {\bibfield  {journal} {\bibinfo  {journal} {Phys. Rev. Fluids}\ }\textbf
  {\bibinfo {volume} {8}},\ \bibinfo {pages} {074802} (\bibinfo {year}
  {2023})}\BibitemShut {NoStop}%
\bibitem [{\citenamefont {Traverso}\ \emph {et~al.}(2023)\citenamefont
  {Traverso}, \citenamefont {Abadie}, \citenamefont {Matar},\ and\
  \citenamefont {Magri}}]{traverso2023data}%
  \BibitemOpen
  \bibfield  {author} {\bibinfo {author} {\bibfnamefont {T.}~\bibnamefont
  {Traverso}}, \bibinfo {author} {\bibfnamefont {T.}~\bibnamefont {Abadie}},
  \bibinfo {author} {\bibfnamefont {O.~K.}\ \bibnamefont {Matar}},\ and\
  \bibinfo {author} {\bibfnamefont {L.}~\bibnamefont {Magri}},\ }\bibfield
  {title} {\bibinfo {title} {Data-driven modeling for drop size
  distributions},\ }\href@noop {} {\bibfield  {journal} {\bibinfo  {journal}
  {Phys. Rev. Fluids}\ }\textbf {\bibinfo {volume} {8}},\ \bibinfo {pages}
  {104302} (\bibinfo {year} {2023})}\BibitemShut {NoStop}%
\bibitem [{\citenamefont {Balla}\ \emph {et~al.}(2020)\citenamefont {Balla},
  \citenamefont {Tripathi},\ and\ \citenamefont {Sahu}}]{balla2020numerical}%
  \BibitemOpen
  \bibfield  {author} {\bibinfo {author} {\bibfnamefont {M.}~\bibnamefont
  {Balla}}, \bibinfo {author} {\bibfnamefont {M.~K.}\ \bibnamefont
  {Tripathi}},\ and\ \bibinfo {author} {\bibfnamefont {K.~C.}\ \bibnamefont
  {Sahu}},\ }\bibfield  {title} {\bibinfo {title} {A numerical study of a
  hollow water droplet falling in air},\ }\href@noop {} {\bibfield  {journal}
  {\bibinfo  {journal} {Theor. Comput. Fluid Dyn.}\ }\textbf {\bibinfo {volume}
  {34}},\ \bibinfo {pages} {133} (\bibinfo {year} {2020})}\BibitemShut
  {NoStop}%
\bibitem [{\citenamefont {Yau}\ and\ \citenamefont
  {Rogers}(1996)}]{yau1996short}%
  \BibitemOpen
  \bibfield  {author} {\bibinfo {author} {\bibfnamefont {M.~K.}\ \bibnamefont
  {Yau}}\ and\ \bibinfo {author} {\bibfnamefont {R.~R.}\ \bibnamefont
  {Rogers}},\ }\href@noop {} {\emph {\bibinfo {title} {A short course in cloud
  physics}}}\ (\bibinfo  {publisher} {Elsevier},\ \bibinfo {year}
  {1996})\BibitemShut {NoStop}%
\bibitem [{\citenamefont {Villermaux}\ and\ \citenamefont
  {Bossa}(2009)}]{Villermaux2009single}%
  \BibitemOpen
  \bibfield  {author} {\bibinfo {author} {\bibfnamefont {E.}~\bibnamefont
  {Villermaux}}\ and\ \bibinfo {author} {\bibfnamefont {B.}~\bibnamefont
  {Bossa}},\ }\bibfield  {title} {\bibinfo {title} {Single-drop fragmentation
  determines size distribution of raindrops},\ }\href@noop {} {\bibfield
  {journal} {\bibinfo  {journal} {Nat. Phys.}\ }\textbf {\bibinfo {volume}
  {5}},\ \bibinfo {pages} {697} (\bibinfo {year} {2009})}\BibitemShut {NoStop}%
\bibitem [{\citenamefont {Ellis}\ \emph {et~al.}(1997)\citenamefont {Ellis},
  \citenamefont {Tuck},\ and\ \citenamefont {Miller}}]{ellis1997effect}%
  \BibitemOpen
  \bibfield  {author} {\bibinfo {author} {\bibfnamefont {M.~C.~B.}\
  \bibnamefont {Ellis}}, \bibinfo {author} {\bibfnamefont {C.~R.}\ \bibnamefont
  {Tuck}},\ and\ \bibinfo {author} {\bibfnamefont {P.~C.~H.}\ \bibnamefont
  {Miller}},\ }\bibfield  {title} {\bibinfo {title} {The effect of some
  adjuvants on sprays produced by agricultural flat fan nozzles},\ }\href@noop
  {} {\bibfield  {journal} {\bibinfo  {journal} {Crop Protection}\ }\textbf
  {\bibinfo {volume} {16}},\ \bibinfo {pages} {41} (\bibinfo {year}
  {1997})}\BibitemShut {NoStop}%
\bibitem [{\citenamefont {Lefebvre}\ and\ \citenamefont
  {McDonell}(2017)}]{lefebvre2017atomization}%
  \BibitemOpen
  \bibfield  {author} {\bibinfo {author} {\bibfnamefont {A.~H.}\ \bibnamefont
  {Lefebvre}}\ and\ \bibinfo {author} {\bibfnamefont {V.~G.}\ \bibnamefont
  {McDonell}},\ }\href@noop {} {\emph {\bibinfo {title} {Atomization and
  sprays}}}\ (\bibinfo  {publisher} {CRC press},\ \bibinfo {year}
  {2017})\BibitemShut {NoStop}%
\bibitem [{\citenamefont {Pilch}\ and\ \citenamefont
  {Erdman}(1987)}]{pilch1987use}%
  \BibitemOpen
  \bibfield  {author} {\bibinfo {author} {\bibfnamefont {M.}~\bibnamefont
  {Pilch}}\ and\ \bibinfo {author} {\bibfnamefont {C.~A.}\ \bibnamefont
  {Erdman}},\ }\bibfield  {title} {\bibinfo {title} {Use of breakup time data
  and velocity history data to predict the maximum size of stable fragments for
  acceleration-induced breakup of a liquid drop},\ }\href@noop {} {\bibfield
  {journal} {\bibinfo  {journal} {Int. J. Multiphase Flow}\ }\textbf {\bibinfo
  {volume} {13}},\ \bibinfo {pages} {741} (\bibinfo {year} {1987})}\BibitemShut
  {NoStop}%
\bibitem [{\citenamefont {Guildenbecher}\ \emph {et~al.}(2009)\citenamefont
  {Guildenbecher}, \citenamefont {L{\'o}pez-Rivera},\ and\ \citenamefont
  {Sojka}}]{guildenbecher2009secondary}%
  \BibitemOpen
  \bibfield  {author} {\bibinfo {author} {\bibfnamefont {D.~R.}\ \bibnamefont
  {Guildenbecher}}, \bibinfo {author} {\bibfnamefont {C.}~\bibnamefont
  {L{\'o}pez-Rivera}},\ and\ \bibinfo {author} {\bibfnamefont {P.~E.}\
  \bibnamefont {Sojka}},\ }\bibfield  {title} {\bibinfo {title} {Secondary
  atomization},\ }\href@noop {} {\bibfield  {journal} {\bibinfo  {journal}
  {Exp. Fluids}\ }\textbf {\bibinfo {volume} {46}},\ \bibinfo {pages} {371}
  (\bibinfo {year} {2009})}\BibitemShut {NoStop}%
\bibitem [{\citenamefont {Suryaprakash}\ and\ \citenamefont
  {Tomar}(2019)}]{suryaprakash2019secondary}%
  \BibitemOpen
  \bibfield  {author} {\bibinfo {author} {\bibfnamefont {R.}~\bibnamefont
  {Suryaprakash}}\ and\ \bibinfo {author} {\bibfnamefont {G.}~\bibnamefont
  {Tomar}},\ }\bibfield  {title} {\bibinfo {title} {Secondary breakup of
  drops},\ }\href@noop {} {\bibfield  {journal} {\bibinfo  {journal} {J. Indian
  Inst. Sci.}\ }\textbf {\bibinfo {volume} {99}},\ \bibinfo {pages} {77}
  (\bibinfo {year} {2019})}\BibitemShut {NoStop}%
\bibitem [{\citenamefont {Soni}\ \emph {et~al.}(2020)\citenamefont {Soni},
  \citenamefont {Kirar}, \citenamefont {Kolhe},\ and\ \citenamefont
  {Sahu}}]{soni2020deformation}%
  \BibitemOpen
  \bibfield  {author} {\bibinfo {author} {\bibfnamefont {S.~K.}\ \bibnamefont
  {Soni}}, \bibinfo {author} {\bibfnamefont {P.~K.}\ \bibnamefont {Kirar}},
  \bibinfo {author} {\bibfnamefont {P.}~\bibnamefont {Kolhe}},\ and\ \bibinfo
  {author} {\bibfnamefont {K.~C.}\ \bibnamefont {Sahu}},\ }\bibfield  {title}
  {\bibinfo {title} {Deformation and breakup of droplets in an oblique
  continuous air stream},\ }\href@noop {} {\bibfield  {journal} {\bibinfo
  {journal} {Int. J. Multiphase Flow}\ }\textbf {\bibinfo {volume} {122}},\
  \bibinfo {pages} {103141} (\bibinfo {year} {2020})}\BibitemShut {NoStop}%
\bibitem [{\citenamefont {Kulkarni}\ \emph {et~al.}(2023)\citenamefont
  {Kulkarni}, \citenamefont {Shirdade}, \citenamefont {Rodrigues},
  \citenamefont {Radhakrishna},\ and\ \citenamefont
  {Sojka}}]{kulkarni2023interdependence}%
  \BibitemOpen
  \bibfield  {author} {\bibinfo {author} {\bibfnamefont {V.}~\bibnamefont
  {Kulkarni}}, \bibinfo {author} {\bibfnamefont {N.}~\bibnamefont {Shirdade}},
  \bibinfo {author} {\bibfnamefont {N.}~\bibnamefont {Rodrigues}}, \bibinfo
  {author} {\bibfnamefont {V.}~\bibnamefont {Radhakrishna}},\ and\ \bibinfo
  {author} {\bibfnamefont {P.~E.}\ \bibnamefont {Sojka}},\ }\bibfield  {title}
  {\bibinfo {title} {On interdependence of instabilities and average drop sizes
  in bag breakup},\ }\href@noop {} {\bibfield  {journal} {\bibinfo  {journal}
  {Appl. Phys. Lett.}\ }\textbf {\bibinfo {volume} {123}} (\bibinfo {year}
  {2023})}\BibitemShut {NoStop}%
\bibitem [{\citenamefont {Taylor}(1963)}]{taylor1963shape}%
  \BibitemOpen
  \bibfield  {author} {\bibinfo {author} {\bibfnamefont {G.~I.}\ \bibnamefont
  {Taylor}},\ }\bibfield  {title} {\bibinfo {title} {The shape and acceleration
  of a drop in a high speed air stream},\ }\href@noop {} {\bibfield  {journal}
  {\bibinfo  {journal} {The Scientific Papers of G. I. Taylor}\ }\textbf
  {\bibinfo {volume} {3}},\ \bibinfo {pages} {457} (\bibinfo {year}
  {1963})}\BibitemShut {NoStop}%
\bibitem [{\citenamefont {Jackiw}\ and\ \citenamefont
  {Ashgriz}(2021)}]{jackiw2021aerodynamic}%
  \BibitemOpen
  \bibfield  {author} {\bibinfo {author} {\bibfnamefont {I.~M.}\ \bibnamefont
  {Jackiw}}\ and\ \bibinfo {author} {\bibfnamefont {N.}~\bibnamefont
  {Ashgriz}},\ }\bibfield  {title} {\bibinfo {title} {On aerodynamic droplet
  breakup},\ }\href@noop {} {\bibfield  {journal} {\bibinfo  {journal} {J.
  Fluid Mech.}\ }\textbf {\bibinfo {volume} {913}},\ \bibinfo {pages} {A33}
  (\bibinfo {year} {2021})}\BibitemShut {NoStop}%
\bibitem [{\citenamefont {Jackiw}\ and\ \citenamefont
  {Ashgriz}(2022)}]{jackiw2022prediction}%
  \BibitemOpen
  \bibfield  {author} {\bibinfo {author} {\bibfnamefont {I.~M.}\ \bibnamefont
  {Jackiw}}\ and\ \bibinfo {author} {\bibfnamefont {N.}~\bibnamefont
  {Ashgriz}},\ }\bibfield  {title} {\bibinfo {title} {Prediction of the droplet
  size distribution in aerodynamic droplet breakup},\ }\href@noop {} {\bibfield
   {journal} {\bibinfo  {journal} {J. Fluid Mech.}\ }\textbf {\bibinfo {volume}
  {940}},\ \bibinfo {pages} {A17} (\bibinfo {year} {2022})}\BibitemShut
  {NoStop}%
\bibitem [{\citenamefont {Kulkarni}\ and\ \citenamefont
  {Sojka}(2014)}]{kulkarni2014bag}%
  \BibitemOpen
  \bibfield  {author} {\bibinfo {author} {\bibfnamefont {V.}~\bibnamefont
  {Kulkarni}}\ and\ \bibinfo {author} {\bibfnamefont {P.~E.}\ \bibnamefont
  {Sojka}},\ }\bibfield  {title} {\bibinfo {title} {Bag breakup of low
  viscosity drops in the presence of a continuous air jet},\ }\href@noop {}
  {\bibfield  {journal} {\bibinfo  {journal} {Phys. Fluids}\ }\textbf {\bibinfo
  {volume} {26}},\ \bibinfo {pages} {072103} (\bibinfo {year}
  {2014})}\BibitemShut {NoStop}%
\bibitem [{\citenamefont {Cao}\ \emph {et~al.}(2007)\citenamefont {Cao},
  \citenamefont {Sun}, \citenamefont {Li}, \citenamefont {Liu},\ and\
  \citenamefont {Yu}}]{cao2007new}%
  \BibitemOpen
  \bibfield  {author} {\bibinfo {author} {\bibfnamefont {X.~K.}\ \bibnamefont
  {Cao}}, \bibinfo {author} {\bibfnamefont {Z.~G.}\ \bibnamefont {Sun}},
  \bibinfo {author} {\bibfnamefont {W.~F.}\ \bibnamefont {Li}}, \bibinfo
  {author} {\bibfnamefont {H.~F.}\ \bibnamefont {Liu}},\ and\ \bibinfo {author}
  {\bibfnamefont {Z.~H.}\ \bibnamefont {Yu}},\ }\bibfield  {title} {\bibinfo
  {title} {A new breakup regime of liquid drops identified in a continuous and
  uniform air jet flow},\ }\href@noop {} {\bibfield  {journal} {\bibinfo
  {journal} {Phys. Fluids}\ }\textbf {\bibinfo {volume} {19}},\ \bibinfo
  {pages} {057103} (\bibinfo {year} {2007})}\BibitemShut {NoStop}%
\bibitem [{\citenamefont {Ade}\ \emph {et~al.}(2023{\natexlab{a}})\citenamefont
  {Ade}, \citenamefont {Chandrala},\ and\ \citenamefont {Sahu}}]{ade2023size}%
  \BibitemOpen
  \bibfield  {author} {\bibinfo {author} {\bibfnamefont {S.~S.}\ \bibnamefont
  {Ade}}, \bibinfo {author} {\bibfnamefont {L.~D.}\ \bibnamefont {Chandrala}},\
  and\ \bibinfo {author} {\bibfnamefont {K.~C.}\ \bibnamefont {Sahu}},\
  }\bibfield  {title} {\bibinfo {title} {Size distribution of a drop undergoing
  breakup at moderate weber numbers},\ }\href@noop {} {\bibfield  {journal}
  {\bibinfo  {journal} {J. Fluid Mech.}\ }\textbf {\bibinfo {volume} {959}},\
  \bibinfo {pages} {A38} (\bibinfo {year} {2023}{\natexlab{a}})}\BibitemShut
  {NoStop}%
\bibitem [{\citenamefont {Joshi}\ and\ \citenamefont
  {Anand}(2022)}]{joshi2022droplet}%
  \BibitemOpen
  \bibfield  {author} {\bibinfo {author} {\bibfnamefont {S.}~\bibnamefont
  {Joshi}}\ and\ \bibinfo {author} {\bibfnamefont {T.~N.~C.}\ \bibnamefont
  {Anand}},\ }\bibfield  {title} {\bibinfo {title} {Droplet deformation in
  secondary breakup: Transformation from a sphere to a disk-like structure},\
  }\href@noop {} {\bibfield  {journal} {\bibinfo  {journal} {Int. J. Multiphase
  Flow}\ }\textbf {\bibinfo {volume} {146}},\ \bibinfo {pages} {103850}
  (\bibinfo {year} {2022})}\BibitemShut {NoStop}%
\bibitem [{\citenamefont {Boggavarapu}\ \emph {et~al.}(2021)\citenamefont
  {Boggavarapu}, \citenamefont {Ramesh}, \citenamefont {Avulapati},\ and\
  \citenamefont {Ravikrishna}}]{boggavarapu2021secondary}%
  \BibitemOpen
  \bibfield  {author} {\bibinfo {author} {\bibfnamefont {P.}~\bibnamefont
  {Boggavarapu}}, \bibinfo {author} {\bibfnamefont {S.~P.}\ \bibnamefont
  {Ramesh}}, \bibinfo {author} {\bibfnamefont {M.~M.}\ \bibnamefont
  {Avulapati}},\ and\ \bibinfo {author} {\bibfnamefont {R.~V.}\ \bibnamefont
  {Ravikrishna}},\ }\bibfield  {title} {\bibinfo {title} {Secondary breakup of
  water and surrogate fuels: Breakup modes and resultant droplet sizes},\
  }\href@noop {} {\bibfield  {journal} {\bibinfo  {journal} {Int. J. Multiphase
  Flow}\ }\textbf {\bibinfo {volume} {145}},\ \bibinfo {pages} {103816}
  (\bibinfo {year} {2021})}\BibitemShut {NoStop}%
\bibitem [{\citenamefont {Kirar}\ \emph {et~al.}(2022)\citenamefont {Kirar},
  \citenamefont {Soni}, \citenamefont {Kolhe},\ and\ \citenamefont
  {Sahu}}]{kirar2022experimental}%
  \BibitemOpen
  \bibfield  {author} {\bibinfo {author} {\bibfnamefont {P.~K.}\ \bibnamefont
  {Kirar}}, \bibinfo {author} {\bibfnamefont {S.~K.}\ \bibnamefont {Soni}},
  \bibinfo {author} {\bibfnamefont {P.~S.}\ \bibnamefont {Kolhe}},\ and\
  \bibinfo {author} {\bibfnamefont {K.~C.}\ \bibnamefont {Sahu}},\ }\bibfield
  {title} {\bibinfo {title} {An experimental investigation of droplet
  morphology in swirl flow},\ }\href@noop {} {\bibfield  {journal} {\bibinfo
  {journal} {J. Fluid Mech.}\ }\textbf {\bibinfo {volume} {938}},\ \bibinfo
  {pages} {A6} (\bibinfo {year} {2022})}\BibitemShut {NoStop}%
\bibitem [{\citenamefont {Ade}\ \emph {et~al.}(2023{\natexlab{b}})\citenamefont
  {Ade}, \citenamefont {Kirar}, \citenamefont {Chandrala},\ and\ \citenamefont
  {Sahu}}]{ade2022droplet}%
  \BibitemOpen
  \bibfield  {author} {\bibinfo {author} {\bibfnamefont {S.~S.}\ \bibnamefont
  {Ade}}, \bibinfo {author} {\bibfnamefont {P.~K.}\ \bibnamefont {Kirar}},
  \bibinfo {author} {\bibfnamefont {L.~D.}\ \bibnamefont {Chandrala}},\ and\
  \bibinfo {author} {\bibfnamefont {K.~C.}\ \bibnamefont {Sahu}},\ }\bibfield
  {title} {\bibinfo {title} {Droplet size distribution in a swirl airstream
  using in-line holography technique},\ }\href@noop {} {\bibfield  {journal}
  {\bibinfo  {journal} {J. Fluid Mech.}\ }\textbf {\bibinfo {volume} {954}},\
  \bibinfo {pages} {A39} (\bibinfo {year} {2023}{\natexlab{b}})}\BibitemShut
  {NoStop}%
\bibitem [{\citenamefont {Tropea}(2011)}]{tropea2011optical}%
  \BibitemOpen
  \bibfield  {author} {\bibinfo {author} {\bibfnamefont {C.}~\bibnamefont
  {Tropea}},\ }\bibfield  {title} {\bibinfo {title} {Optical particle
  characterization in flows},\ }\href@noop {} {\bibfield  {journal} {\bibinfo
  {journal} {Ann. Rev. Fluid Mech.}\ }\textbf {\bibinfo {volume} {43}},\
  \bibinfo {pages} {399} (\bibinfo {year} {2011})}\BibitemShut {NoStop}%
\bibitem [{\citenamefont {Guildenbecher}\ \emph {et~al.}(2017)\citenamefont
  {Guildenbecher}, \citenamefont {Gao}, \citenamefont {Chen},\ and\
  \citenamefont {Sojka}}]{guildenbecher2017characterization}%
  \BibitemOpen
  \bibfield  {author} {\bibinfo {author} {\bibfnamefont {D.~R.}\ \bibnamefont
  {Guildenbecher}}, \bibinfo {author} {\bibfnamefont {J.}~\bibnamefont {Gao}},
  \bibinfo {author} {\bibfnamefont {J.}~\bibnamefont {Chen}},\ and\ \bibinfo
  {author} {\bibfnamefont {P.~E.}\ \bibnamefont {Sojka}},\ }\bibfield  {title}
  {\bibinfo {title} {Characterization of drop aerodynamic fragmentation in the
  bag and sheet-thinning regimes by crossed-beam, two-view, digital in-line
  holography},\ }\href@noop {} {\bibfield  {journal} {\bibinfo  {journal} {Int.
  J. Multiphase Flow}\ }\textbf {\bibinfo {volume} {94}},\ \bibinfo {pages}
  {107} (\bibinfo {year} {2017})}\BibitemShut {NoStop}%
\bibitem [{\citenamefont {Shao}\ \emph {et~al.}(2020)\citenamefont {Shao},
  \citenamefont {Mallery}, \citenamefont {Kumar},\ and\ \citenamefont
  {Hong}}]{shao2020machine}%
  \BibitemOpen
  \bibfield  {author} {\bibinfo {author} {\bibfnamefont {S.}~\bibnamefont
  {Shao}}, \bibinfo {author} {\bibfnamefont {K.}~\bibnamefont {Mallery}},
  \bibinfo {author} {\bibfnamefont {S.~S.}\ \bibnamefont {Kumar}},\ and\
  \bibinfo {author} {\bibfnamefont {J.}~\bibnamefont {Hong}},\ }\bibfield
  {title} {\bibinfo {title} {Machine learning holography for {3D} particle
  field imaging},\ }\href@noop {} {\bibfield  {journal} {\bibinfo  {journal}
  {Optics Express}\ }\textbf {\bibinfo {volume} {28}},\ \bibinfo {pages} {2987}
  (\bibinfo {year} {2020})}\BibitemShut {NoStop}%
\bibitem [{\citenamefont {Radhakrishna}\ \emph {et~al.}(2021)\citenamefont
  {Radhakrishna}, \citenamefont {Shang}, \citenamefont {Yao}, \citenamefont
  {Chen},\ and\ \citenamefont {Sojka}}]{radhakrishna2021experimental}%
  \BibitemOpen
  \bibfield  {author} {\bibinfo {author} {\bibfnamefont {V.}~\bibnamefont
  {Radhakrishna}}, \bibinfo {author} {\bibfnamefont {W.}~\bibnamefont {Shang}},
  \bibinfo {author} {\bibfnamefont {L.}~\bibnamefont {Yao}}, \bibinfo {author}
  {\bibfnamefont {J.}~\bibnamefont {Chen}},\ and\ \bibinfo {author}
  {\bibfnamefont {P.~E.}\ \bibnamefont {Sojka}},\ }\bibfield  {title} {\bibinfo
  {title} {Experimental characterization of secondary atomization at high
  ohnesorge numbers},\ }\href@noop {} {\bibfield  {journal} {\bibinfo
  {journal} {Int. J. Multiphase Flow}\ }\textbf {\bibinfo {volume} {138}},\
  \bibinfo {pages} {103591} (\bibinfo {year} {2021})}\BibitemShut {NoStop}%
\bibitem [{\citenamefont {Essa{\"\i}di}\ \emph {et~al.}(2021)\citenamefont
  {Essa{\"\i}di}, \citenamefont {Lauret}, \citenamefont {Heymes}, \citenamefont
  {Aprin},\ and\ \citenamefont {Slangen}}]{essaidi2021aerodynamic}%
  \BibitemOpen
  \bibfield  {author} {\bibinfo {author} {\bibfnamefont {Z.}~\bibnamefont
  {Essa{\"\i}di}}, \bibinfo {author} {\bibfnamefont {P.}~\bibnamefont
  {Lauret}}, \bibinfo {author} {\bibfnamefont {F.}~\bibnamefont {Heymes}},
  \bibinfo {author} {\bibfnamefont {L.}~\bibnamefont {Aprin}},\ and\ \bibinfo
  {author} {\bibfnamefont {P.}~\bibnamefont {Slangen}},\ }\bibfield  {title}
  {\bibinfo {title} {Aerodynamic fragmentation of water, ethanol and
  polyethylene glycol droplets investigated by high-speed in-line digital
  holography},\ }\href@noop {} {\bibfield  {journal} {\bibinfo  {journal}
  {Optical Materials}\ }\textbf {\bibinfo {volume} {122}},\ \bibinfo {pages}
  {111747} (\bibinfo {year} {2021})}\BibitemShut {NoStop}%
\bibitem [{\citenamefont {Li}\ \emph {et~al.}(2022)\citenamefont {Li},
  \citenamefont {Shen}, \citenamefont {Liu}, \citenamefont {Zhao},
  \citenamefont {Li},\ and\ \citenamefont {Tang}}]{li2022secondary}%
  \BibitemOpen
  \bibfield  {author} {\bibinfo {author} {\bibfnamefont {J.}~\bibnamefont
  {Li}}, \bibinfo {author} {\bibfnamefont {S.}~\bibnamefont {Shen}}, \bibinfo
  {author} {\bibfnamefont {J.}~\bibnamefont {Liu}}, \bibinfo {author}
  {\bibfnamefont {Y.}~\bibnamefont {Zhao}}, \bibinfo {author} {\bibfnamefont
  {S.}~\bibnamefont {Li}},\ and\ \bibinfo {author} {\bibfnamefont
  {C.}~\bibnamefont {Tang}},\ }\bibfield  {title} {\bibinfo {title} {Secondary
  droplet size distribution upon breakup of a sub-milimeter droplet in high
  speed cross flow},\ }\href@noop {} {\bibfield  {journal} {\bibinfo  {journal}
  {Int. J. Multiphase Flow}\ }\textbf {\bibinfo {volume} {148}},\ \bibinfo
  {pages} {103943} (\bibinfo {year} {2022})}\BibitemShut {NoStop}%
\bibitem [{\citenamefont {Montero-Mart{\'\i}nez}\ \emph
  {et~al.}(2009)\citenamefont {Montero-Mart{\'\i}nez}, \citenamefont
  {Kostinski}, \citenamefont {Shaw},\ and\ \citenamefont
  {Garc{\'\i}a-Garc{\'\i}a}}]{montero2009all}%
  \BibitemOpen
  \bibfield  {author} {\bibinfo {author} {\bibfnamefont {G.}~\bibnamefont
  {Montero-Mart{\'\i}nez}}, \bibinfo {author} {\bibfnamefont {A.~B.}\
  \bibnamefont {Kostinski}}, \bibinfo {author} {\bibfnamefont {R.~A.}\
  \bibnamefont {Shaw}},\ and\ \bibinfo {author} {\bibfnamefont
  {F.}~\bibnamefont {Garc{\'\i}a-Garc{\'\i}a}},\ }\bibfield  {title} {\bibinfo
  {title} {Do all raindrops fall at terminal speed?},\ }\href@noop {}
  {\bibfield  {journal} {\bibinfo  {journal} {Geophys. Res. Lett.}\ }\textbf
  {\bibinfo {volume} {36}} (\bibinfo {year} {2009})}\BibitemShut {NoStop}%
\bibitem [{\citenamefont {Ade}\ \emph {et~al.}(2024)\citenamefont {Ade},
  \citenamefont {Gupta}, \citenamefont {Chandrala},\ and\ \citenamefont
  {Sahu}}]{ade2024application}%
  \BibitemOpen
  \bibfield  {author} {\bibinfo {author} {\bibfnamefont {S.~S.}\ \bibnamefont
  {Ade}}, \bibinfo {author} {\bibfnamefont {D.}~\bibnamefont {Gupta}}, \bibinfo
  {author} {\bibfnamefont {L.~D.}\ \bibnamefont {Chandrala}},\ and\ \bibinfo
  {author} {\bibfnamefont {K.~C.}\ \bibnamefont {Sahu}},\ }\bibfield  {title}
  {\bibinfo {title} {Application of deep learning and inline holography to
  estimate the droplet size distribution},\ }\href@noop {} {\bibfield
  {journal} {\bibinfo  {journal} {Int. J. Multiphase Flow}\ }\textbf {\bibinfo
  {volume} {177}},\ \bibinfo {pages} {104853} (\bibinfo {year}
  {2024})}\BibitemShut {NoStop}%
\bibitem [{\citenamefont {Rayleigh}(1879)}]{rayleigh1879capillary}%
  \BibitemOpen
  \bibfield  {author} {\bibinfo {author} {\bibfnamefont {L.}~\bibnamefont
  {Rayleigh}},\ }\bibfield  {title} {\bibinfo {title} {On the capillary
  phenomena of jets},\ }\href@noop {} {\bibfield  {journal} {\bibinfo
  {journal} {Proc. R. Soc. London}\ }\textbf {\bibinfo {volume} {29}},\
  \bibinfo {pages} {71} (\bibinfo {year} {1879})}\BibitemShut {NoStop}%
\bibitem [{\citenamefont {Nelson}\ and\ \citenamefont
  {Gokhale}(1972)}]{nelson1972oscillation}%
  \BibitemOpen
  \bibfield  {author} {\bibinfo {author} {\bibfnamefont {A.~R.}\ \bibnamefont
  {Nelson}}\ and\ \bibinfo {author} {\bibfnamefont {N.~R.}\ \bibnamefont
  {Gokhale}},\ }\bibfield  {title} {\bibinfo {title} {Oscillation frequencies
  of freely suspended water drops},\ }\href@noop {} {\bibfield  {journal}
  {\bibinfo  {journal} {J. Geophys. Res.}\ }\textbf {\bibinfo {volume} {77}},\
  \bibinfo {pages} {2724} (\bibinfo {year} {1972})}\BibitemShut {NoStop}%
\bibitem [{\citenamefont {Agrawal}\ \emph {et~al.}(2017)\citenamefont
  {Agrawal}, \citenamefont {Premlata}, \citenamefont {Tripathi}, \citenamefont
  {Karri},\ and\ \citenamefont {Sahu}}]{agrawal2017nonspherical}%
  \BibitemOpen
  \bibfield  {author} {\bibinfo {author} {\bibfnamefont {M.}~\bibnamefont
  {Agrawal}}, \bibinfo {author} {\bibfnamefont {A.~R.}\ \bibnamefont
  {Premlata}}, \bibinfo {author} {\bibfnamefont {M.~K.}\ \bibnamefont
  {Tripathi}}, \bibinfo {author} {\bibfnamefont {B.}~\bibnamefont {Karri}},\
  and\ \bibinfo {author} {\bibfnamefont {K.~C.}\ \bibnamefont {Sahu}},\
  }\bibfield  {title} {\bibinfo {title} {Nonspherical liquid droplet falling in
  air},\ }\href@noop {} {\bibfield  {journal} {\bibinfo  {journal} {Phys. Rev.
  E.}\ }\textbf {\bibinfo {volume} {95}},\ \bibinfo {pages} {033111} (\bibinfo
  {year} {2017})}\BibitemShut {NoStop}%
\bibitem [{\citenamefont {Balla}\ \emph {et~al.}(2019)\citenamefont {Balla},
  \citenamefont {Tripathi},\ and\ \citenamefont {Sahu}}]{balla2019shape}%
  \BibitemOpen
  \bibfield  {author} {\bibinfo {author} {\bibfnamefont {M.}~\bibnamefont
  {Balla}}, \bibinfo {author} {\bibfnamefont {M.~K.}\ \bibnamefont
  {Tripathi}},\ and\ \bibinfo {author} {\bibfnamefont {K.~C.}\ \bibnamefont
  {Sahu}},\ }\bibfield  {title} {\bibinfo {title} {Shape oscillations of a
  nonspherical water droplet},\ }\href@noop {} {\bibfield  {journal} {\bibinfo
  {journal} {Phys. Rev. E.}\ }\textbf {\bibinfo {volume} {99}},\ \bibinfo
  {pages} {023107} (\bibinfo {year} {2019})}\BibitemShut {NoStop}%
\bibitem [{\citenamefont {Agrawal}\ \emph {et~al.}(2020)\citenamefont
  {Agrawal}, \citenamefont {Katiyar}, \citenamefont {Karri},\ and\
  \citenamefont {Sahu}}]{agrawal2020experimental}%
  \BibitemOpen
  \bibfield  {author} {\bibinfo {author} {\bibfnamefont {M.}~\bibnamefont
  {Agrawal}}, \bibinfo {author} {\bibfnamefont {R.~K.}\ \bibnamefont
  {Katiyar}}, \bibinfo {author} {\bibfnamefont {B.}~\bibnamefont {Karri}},\
  and\ \bibinfo {author} {\bibfnamefont {K.~C.}\ \bibnamefont {Sahu}},\
  }\bibfield  {title} {\bibinfo {title} {Experimental investigation of a
  nonspherical water droplet falling in air},\ }\href@noop {} {\bibfield
  {journal} {\bibinfo  {journal} {Phys. Fluids}\ }\textbf {\bibinfo {volume}
  {32}},\ \bibinfo {pages} {112105} (\bibinfo {year} {2020})}\BibitemShut
  {NoStop}%
\bibitem [{\citenamefont {Zhao}\ \emph {et~al.}(2010)\citenamefont {Zhao},
  \citenamefont {Liu}, \citenamefont {Li},\ and\ \citenamefont
  {Xu}}]{zhao2010morphological}%
  \BibitemOpen
  \bibfield  {author} {\bibinfo {author} {\bibfnamefont {H.}~\bibnamefont
  {Zhao}}, \bibinfo {author} {\bibfnamefont {H.~F.}\ \bibnamefont {Liu}},
  \bibinfo {author} {\bibfnamefont {W.~F.}\ \bibnamefont {Li}},\ and\ \bibinfo
  {author} {\bibfnamefont {J.~L.}\ \bibnamefont {Xu}},\ }\bibfield  {title}
  {\bibinfo {title} {Morphological classification of low viscosity drop bag
  breakup in a continuous air jet stream},\ }\href@noop {} {\bibfield
  {journal} {\bibinfo  {journal} {Phys. Fluids}\ }\textbf {\bibinfo {volume}
  {22}},\ \bibinfo {pages} {114103} (\bibinfo {year} {2010})}\BibitemShut
  {NoStop}%
\bibitem [{\citenamefont {Wang}\ \emph {et~al.}(2018)\citenamefont {Wang},
  \citenamefont {Dandekar}, \citenamefont {Bustos}, \citenamefont {Poulain},\
  and\ \citenamefont {Bourouiba}}]{wang2018universal}%
  \BibitemOpen
  \bibfield  {author} {\bibinfo {author} {\bibfnamefont {Y.}~\bibnamefont
  {Wang}}, \bibinfo {author} {\bibfnamefont {R.}~\bibnamefont {Dandekar}},
  \bibinfo {author} {\bibfnamefont {N.}~\bibnamefont {Bustos}}, \bibinfo
  {author} {\bibfnamefont {S.}~\bibnamefont {Poulain}},\ and\ \bibinfo {author}
  {\bibfnamefont {L.}~\bibnamefont {Bourouiba}},\ }\bibfield  {title} {\bibinfo
  {title} {Universal rim thickness in unsteady sheet fragmentation},\
  }\href@noop {} {\bibfield  {journal} {\bibinfo  {journal} {Physical review
  letters}\ }\textbf {\bibinfo {volume} {120}},\ \bibinfo {pages} {204503}
  (\bibinfo {year} {2018})}\BibitemShut {NoStop}%
\bibitem [{\citenamefont {Keshavarz}\ \emph {et~al.}(2020)\citenamefont
  {Keshavarz}, \citenamefont {Houze}, \citenamefont {Moore}, \citenamefont
  {Koerner},\ and\ \citenamefont {McKinley}}]{keshavarz2020rotary}%
  \BibitemOpen
  \bibfield  {author} {\bibinfo {author} {\bibfnamefont {B.}~\bibnamefont
  {Keshavarz}}, \bibinfo {author} {\bibfnamefont {E.~C.}\ \bibnamefont
  {Houze}}, \bibinfo {author} {\bibfnamefont {J.~R.}\ \bibnamefont {Moore}},
  \bibinfo {author} {\bibfnamefont {M.~R.}\ \bibnamefont {Koerner}},\ and\
  \bibinfo {author} {\bibfnamefont {G.~H.}\ \bibnamefont {McKinley}},\
  }\bibfield  {title} {\bibinfo {title} {Rotary atomization of newtonian and
  viscoelastic liquids},\ }\href@noop {} {\bibfield  {journal} {\bibinfo
  {journal} {Phys. Rev. Fluids}\ }\textbf {\bibinfo {volume} {5}},\ \bibinfo
  {pages} {033601} (\bibinfo {year} {2020})}\BibitemShut {NoStop}%
\end{thebibliography}

%

\end{document}